\documentclass[iop, tighten]{emulateapj}

\usepackage{amsmath}
\usepackage[usenames, dvipsnames]{color}  
\usepackage{hyperref}
\hypersetup{colorlinks=True, citecolor=blue, linkcolor=Violet}  

\newcommand{\msun}{\ensuremath{\, \rm{M}_\odot}}
\newcommand{\ssec}{\ensuremath{\,\mathrm{s}}}

\newcommand{\erg}{\ensuremath{\,\mathrm{erg}}}
\newcommand{\cm}{\ensuremath{\,\mathrm{cm}}}
\newcommand{\pc}{\ensuremath{\,\mathrm{pc}}}

\newcommand{\ang}{\ensuremath{\,\textrm{\AA}}}

\newcommand{\uflambda}{\ensuremath{\, \erg\ssec^{-1}\cm^{-2}\ang^{-1}}}
\newcommand{\degree}{\ensuremath{^\circ}}

\newcommand{\fuv}{\ensuremath{\mathrm{FUV}}}
\newcommand{\nuv}{\ensuremath{\mathrm{NUV}}}
\newcommand{\acsb}{\ensuremath{\mathrm{F475W}}}
\newcommand{\acsi}{\ensuremath{\mathrm{F814W}}}

\newcommand{\sfh}{\ensuremath{\mathrm{SFH}}}
\newcommand{\sfr}{\ensuremath{\mathrm{SFR}}}
\newcommand{\meansfr}{\ensuremath{\langle\sfr\rangle}}

\newcommand{\met}{\ensuremath{\mathrm{[M/H]}}}
\newcommand{\avsfh}{\ensuremath{A_{\mathrm{V, SFH}}}}
\newcommand{\davsfh}{\ensuremath{dA_\mathrm{V, SFH}}}

\newcommand{\rv}{\ensuremath{R_\mathrm{V}}}

\newcommand{\obs}{\ensuremath{\mathrm{obs}}}
\newcommand{\syn}{\ensuremath{\mathrm{syn}}}

\newcommand{\fuvnuvobs}{\ensuremath{(m_\fuv-m_\nuv)^\obs}}
\newcommand{\fuvnuvsfh}{\ensuremath{(m_\fuv-m_\nuv)^\syn}}

\newcommand{\fuvnuv}{\ensuremath{m_\fuv-m_\nuv}}

\newcommand{\fxobs}{\ensuremath{f^\obs}} 
\newcommand{\ffuvobs}{\ensuremath{f_\fuv^\obs}}
\newcommand{\fnuvobs}{\ensuremath{f_\nuv^\obs}}
\newcommand{\fxsfh}{\ensuremath{f^\syn}}  
\newcommand{\ffuvsfh}{\ensuremath{f_\fuv^\syn}}
\newcommand{\fnuvsfh}{\ensuremath{f_\nuv^\syn}}
\newcommand{\fxsfhz}{\ensuremath{f^{\syn,0}}}  
\newcommand{\ffuvsfhz}{\ensuremath{f_\fuv^{\syn,0}}}

\newcommand{\sfroneh}{\ensuremath{\meansfr_{100}}}

\newcommand{\galex}{\textit{GALEX}}
\newcommand{\ha}{\ensuremath{\mathrm{H}\alpha}}

\newcommand{\logten}{\ensuremath{\log_{10}}}

\newcommand{\sfrunit}{\ensuremath{\msun\;{\rm yr}^{-1}}}
\def\about  {\hbox{$\sim$}}

\newif\ifshownuv
\shownuvtrue

\begin{document}

\title{The Panchromatic Hubble Andromeda Treasury. \textsc{XVII}. Examining Obscured Star Formation with Synthetic Ultraviolet Flux Maps in M31.\footnotemark[*]}

\footnotetext[*]{Based on observations made with the NASA/ESA Hubble Space Telescope, obtained at the Space Telescope Science Institute, which is operated by the Association of Universities for Research in Astronomy, Inc., under NASA contract NAS 5-26555. These observations are associated with program \#12055.}

\author{Alexia R. Lewis\altaffilmark{1,2,3},
Jacob E. Simones\altaffilmark{4},
Benjamin D. Johnson\altaffilmark{5},
Julianne J. Dalcanton\altaffilmark{1},
Evan D. Skillman\altaffilmark{4},
Daniel R. Weisz\altaffilmark{6},
Andrew E. Dolphin\altaffilmark{7},
Benjamin F. Williams\altaffilmark{1},
Eric F. Bell\altaffilmark{8},
Morgan Fouesneau\altaffilmark{9},
Maria Kapala \altaffilmark{10},
Philip Rosenfield\altaffilmark{5},
Andreas Schruba\altaffilmark{11}
}

\altaffiltext{1}{Department of Astronomy, Box 351580, University of Washington, Seattle, WA 98195, USA}
\altaffiltext{2}{Center for Cosmology and AstroParticle Physics, The Ohio State University, Columbus, OH 43210, USA}
\altaffiltext{3}{Department of Astronomy, The Ohio State University, 140 West 18th Avenue, Columbus, OH 43210, USA}
\altaffiltext{4}{Minnesota Institute for Astrophysics, University of Minnesota, 116 Church Street SE, Minneapolis, MN 55455, USA}
\altaffiltext{5}{Harvard-Smithsonian Center for Astrophysics, 60 Garden Street, Cambridge, MA 02138, USA}
\altaffiltext{6}{Department of Astronomy, University of California Berkeley, Berkeley, CA 94720, USA}
\altaffiltext{7}{Raytheon, 1151 E. Hermans Road, Tucson, AZ 85756, USA}
\altaffiltext{8}{Department of Astronomy, University of Michigan, 1085 S. University Ave., Ann Arbor, MI 48109, USA}
\altaffiltext{9}{Max-Planck-Institut f\"ur Astronomie, K\"onigstul 17, D-69117 Heidelberg, Germany}
\altaffiltext{10}{Department of Astronomy, University of Cape Town, Private Bag X3, Rondebosch 7701, South Africa}
\altaffiltext{11}{Max-Planck-Institut f\"ur extraterrestrische Physik, Giessenbachstrasse 1, 85748 Garching, Germany}

\shortauthors{Lewis et al.}

\begin{abstract}

We present synthetic far- and near-ultraviolet (\fuv{} and \nuv{}) maps of M31, both with and without dust reddening. These maps were constructed from spatially-resolved star formation histories (\sfh{s}) derived from optical \textit{Hubble Space Telescope} imaging of resolved stars, taken as part of the Panchromatic Hubble Andromeda Treasury (PHAT) program. We use stellar population synthesis modeling to generate synthetic UV maps with a spatial resolution of \about100 pc (\about24 arcseconds), projected. When reddening is included, these maps reproduce all of the main morphological features in the \galex{} imaging, including rings and large star-forming complexes. The predicted UV flux also agrees well with the observed flux, with median ratios between the modeled and observed flux of $\log_{10}(\ffuvsfh{}/\ffuvobs{}) = 0.03 \pm 0.24$ and $\log_{10}(\fnuvsfh{}/\fnuvobs{}) = -0.03 \pm 0.16$ in the \fuv{} and \nuv{}, respectively. This agreement is particularly impressive given that we used only optical photometry to construct these UV maps. Having verified the synthetic reddened maps, we use the dust-free maps to examine properties of obscured flux and star formation. We compare our dust-free and reddened maps of \fuv{} flux with the observed \galex{} \fuv{} flux and \fuv{} + 24 \micron{} flux to examine the fraction of obscured flux. We find that the maps of synthetic flux require that \about90\% of the \fuv{} flux in M31 is obscured by dust, while the \galex{}-based methods suggest that \about70\% of the \fuv{} flux is absorbed by dust. This 30\% increase in the estimate of the obscured flux is driven by significant differences between the dust-free synthetic \fuv{} flux and that derived when correcting the observed \fuv{} flux for dust absorption with 24 \micron{} emission observations. The difference is further illustrated when we compare the \sfr{s} derived from the \fuv{} + 24 \micron{} flux with the 100 Myr average \sfr{} from the CMD-based \sfh{s}. We find that the 24 \micron{}-corrected \fuv{} flux underestimates the \sfr{} by a factor of 2.3 -- 2.5, depending on the chosen calibration. This discrepancy could be reduced by allowing for variability in the weight applied to the 24 \micron{} data, as has been recently suggested in the literature.

\end{abstract}

\keywords{
    galaxies: evolution --
    galaxies: individual (M31) --
    galaxies: star formation --
    galaxies: stellar content
}

\section{Introduction}
\label{sec:intro}

The star formation rate (\sfr{}) of a galaxy is an extremely import astrophysical quantity, a key component to understanding the detailed evolution of a single galaxy or that of a population of galaxies across cosmic time. To this end, the ultraviolet (UV) is one of the most diagnostically important parts of a galaxy's spectrum. It is often interpreted as a measure of recent star formation, given that stars younger than 300 Myr, including young, massive O and B stars, emit most of their energy in the UV \citep[e.g.,][]{Kennicutt2012a}. Additionally, the UV is essential for tracing star formation across cosmic time. At high redshift, the light observed in the optical is the galaxy's rest frame UV, which has been redshifted to longer wavelengths.

The UV, however, is also one of the most challenging parts of the spectrum to interpret reliably. While it is generally a good measure of recent star formation, the UV is also highly sensitive to dust. UV light is absorbed by dust grains, which re-emit in the infrared. Consequently, UV-only star formation rates (\sfr{s}) underestimate the true \sfr{} of a galaxy, usually by factors of a few in typical disk galaxies \citep[e.g.,][]{Kennicutt1998a, Leroy2012a}, though it can be as much as a factor of 100 or more in ULIRGS \citep[e.g.,][]{Schmitt2006a}. Additionally, stellar models in the UV are difficult to create and are neither well tested nor well-calibrated \citep[e.g.,][]{Pradhan2014a}. Models are particularly problematic for highly evolved low-mass stars, which also emit modestly in the UV \citep[e.g.,][]{Code1969a, OConnell1992a, Dorman1993a, Rosenfield2012a, Johnson2013a}, making the UV flux from older stellar populations particularly difficult to interpret \citep{Conroy2013a}. Finally, UV observations are difficult for low-redshift galaxies because they requires a space or balloon-borne mission, like the \textit{Galaxy Evolution Explorer} \citep[\galex;][]{Martin2005a}, the UVOT camera on \textit{Swift} \citep{Gehrels2004a}, or the UVIS channel on the Wide Field Planetary Camera 3 (WFC3/UVIS) aboard the \textit{Hubble Space Telescope}.

Further complications when interpreting UV observations arise from assumptions on the initial mass function (IMF) and the \sfr{}. Models generally assume that (1) the IMF is fully populated and (2) that the SFR is constant at recent times. The first of these assumptions is likely to hold only over large areas or in small but high star formation surface density regions. The second requirement arise because, at UV wavelengths, up to 90\% of the emission in the far UV (\fuv{}) and near UV (\nuv{}) is from stars that are younger than 100 Myr and 300 Myr, respectively; conversion to a \sfr{} therefore assumes that the \sfr{} has been uniform over that period of time \citep[e.g.,][]{Kennicutt2012a}.

Although a wide range of science is enabled with these assumptions, the circumstances under which they are reliable remain unclear. For example, recent work on the IMF has shown that for sufficiently low SFRs and/or small spatial scales, the high-mass IMF is not fully populated \citep{da-Silva2012a, da-Silva2014a, Krumholz2015a}, and may even systematically vary \citep[e.g.,][]{Meurer2009a}. Similarly, in low-mass galaxies and/or on small spatial scales in large galaxies, star formation histories (\sfh{s}) tend to be bursty as opposed to constant \citep[e.g.,][]{Lee2009a, Weisz2012a}. The consequences of these deviations from the fiducial assumptions are not well-understood and may have a significant impact on our interpretation of the observed flux and consequently on the \sfr{}. 

Given the astrophysical importance of \sfr{} measurements, tests of the above models and assumptions are crucial. A number of studies have examined how interpreting the observed galaxy flux is affected by some of these assumptions \citep[e.g.,][among many others]{Lee2009a, Chomiuk2011a, Weisz2012a, Wilkins2012a, Johnson2013a, Boquien2014a, Simones2014a}. In a recent paper, \citet{Boquien2015a} analyzed the scale dependence of \sfr{} tracers. They found broad agreement between tracers on \about1 kpc scales, presumably because variations between regions of active star formation and diffuse emission (i.e., from older stellar populations) have averaged out. The scale at which this occurs will vary from galaxy to galaxy depending on the intensity of star formation as well as the structure and transparency of the interstellar medium.

\begin{figure}
\centering
\includegraphics[width=\columnwidth]{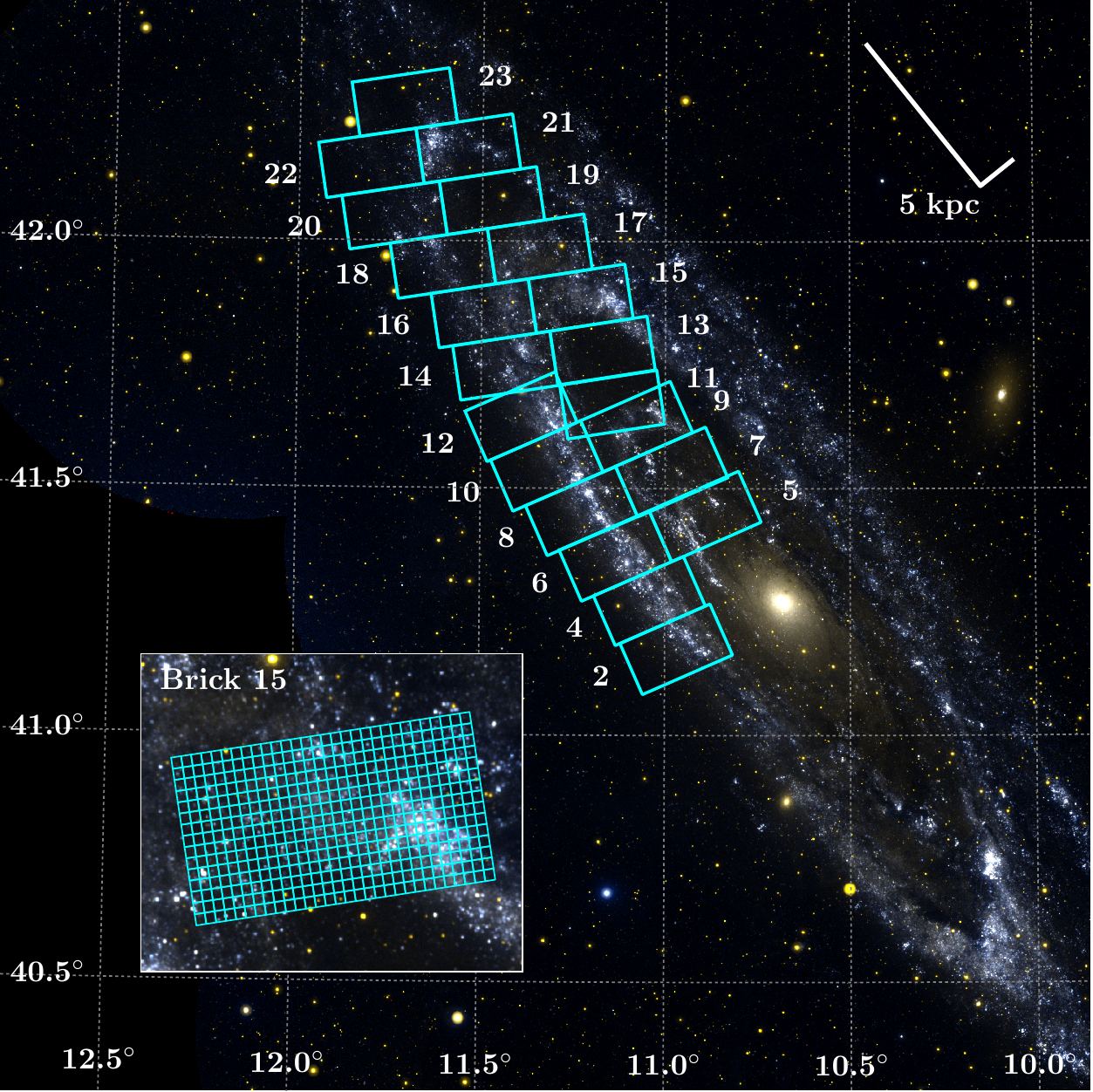}
\caption[PHAT survey map.]{Map of the PHAT survey area. The 21 PHAT bricks analyzed in this study are outlined and numbered. Each brick was divided into 450 regions on a $15 \times 30$ grid, as shown for Brick 15 in the inset panel. The \sfh{s} for each region are presented in \citet{Lewis2015a}. We note that Bricks 1 and 3 (nearest to the center of the galaxy) are not used in the present study because their CMDs are too shallow for reliable \sfh{} determination. In the upper right image we include a scale bar denoting 5 kpc along the major and minor axes of the galaxy. The image is oriented such that north is up and east is to the left.}
\label{fig:map}
\end{figure}

In this paper, we take an alternate approach to \sfr{} analysis that sidesteps some of the above assumptions.
We use \sfh{s} and dust distributions derived from resolved stars and stellar population synthesis models to create maps of synthetic \fuv{} and \nuv{} flux on sub-kpc scales. The \sfh{s} come from \citet{Lewis2015a}, who used optical\footnote{Although PHAT includes data in two UV filters (F275W and F336W), that data is not used in this analysis. Only the brightest main sequence stars have measurements in the UV filters, severely limiting the age range over which the \sfh{} would be reliable. The optical data are the deepest and therefore provide the best leverage on measuring the \sfh{}.} resolved star data, taken as part of the Panchromatic Hubble Andromeda Treasury \citep[PHAT;][]{Dalcanton2012a, Williams2014a} program, to model the spatially-resolved recent \sfh{} of \about9000 regions (\about0.01 sq. kpc in size) in the northeast quadrant of M31 (Figure \ref{fig:map}). Use of these detailed \sfh{s} allows us to relax the constant \sfr{} assumption, and the poisson likelihood sampling in the \sfh{} derivation process implicitly corrects the \sfh{s} for IMF sampling, which is included in the uncertainty estimate. We compare the synthetic flux maps with far- and near-UV \galex{} observations \citep{Morrissey2007a} to analyze the effect of non-constant \sfh{s} on UV flux over a wide range of environments within a single galaxy. This methodology facilitates an end-to-end test verifying the connection between massive star formation, UV flux, and the ability to interpret and/or infer UV populations through analysis of optical stellar populations.

\citet{Simones2014a} initiated this work in M31, using techniques similar to that of \citet{Johnson2013a} to 
model the UV flux in 33 star-forming regions in M31's star-forming ring and compare UV flux-derived \sfr{s} to those measured from optical color-magnitude diagram (CMD)-derived \sfh{s}. They found that the \sfr{s} derived from CMDs were, on average, consistent with those derived from the extinction-corrected \fuv{} flux to within $1\sigma$, and that \about1/3 of the scatter could be attributed to metallicity differences: the flux calibration assumes constant solar metallicity, while the metallicity of the SFHs varies with time. Additionally, \citet{Simones2014a} found that a wide range of \sfh{s} can produce the same amount of \fuv{} flux.

The analysis of resolved stellar populations offers a different way to probe the effects of simplifying assumptions on integrated flux measurements. Unfortunately, tests such as the one performed by \citet{Simones2014a} have been limited in scope, both due to the inability to gather the required resolved star data outside of the Local Group and because the vast majority of galaxies in the nearby universe are dwarf galaxies. \citet{Johnson2013a} explored the connection between the UV and resolved stellar populations in nearby low-mass, low-metallicity dwarf galaxies using \sfh{s} from the ACS Nearby Galaxy Treasury program \citep[ANGST;][]{Dalcanton2009a, Weisz2011a}. They found that fluctuations in the \sfh{} can cause factor of two variations in UV luminosities relative to constant \sfr{} assumptions, and that stars older than 100 Myr can contribute up to 30\% of FUV emission.

Using the \sfh{s} from resolved CMD fitting is just one way to model the UV flux with stellar photometry. Another method is to fit the spectral energy distributions (SEDs) of individual stars \citep[e.g.,][]{Romaniello2002a, Maiz-Apellaniz2004a, Robitaille2007a, Bailer-Jones2011a, Bianchi2012b, Bianchi2012a}. SED fitting is often performed to recover the properties of the star and the surrounding interstellar medium ($T_{\rm eff}$, $\log g$, $Z$, $A_V$). Such fitting, though, ideally produces the star's full spectrum, both reddened and intrinsic, which can then be used to infer the flux at wavelengths that were not observed. The PHAT survey motivated the development of the Bayesian Extinction and Stellar Tool \citep[BEAST;][]{Gordon2016a}, an SED fitter, which could be used to provide and independent check of the results presented in this paper. SED fitting will, however, miss any stars below the detection limit because it requires that they be detected in multiple filters. In contrast, the CMD modeling used in this paper includes the UV flux for the entire stellar population.

Making this measurement in a large, Milky Way-like galaxy is important for interpreting the observations of massive galaxies that dominate star-forming galaxy samples. In the nearby universe, we are limited to only a few large spiral galaxies with sufficiently well-resolved populations. Of these, M31 data has the widest wavelength coverage, the best resolved star data, and the highest resolution. M31 is a more massive and metal-rich system than nearby dwarf galaxies, providing an important laboratory for testing the assumptions made when using integrated UV light. While the method for studying UV populations is the same as that in lower-mass systems, M31's environment is critical because it is similar to that in which most stars are formed.

This paper is organized as follows. In Section \ref{sec:data}, we describe the \sfh{} dataset and the \galex{} \fuv{} and \nuv{} images used, and the production of the synthetic flux maps. Section \ref{sec:map_creation} describes the creation of the UV maps, including the methodology used to model the UV fluxes and the manner in which we assembled the maps. We present the maps in Section \ref{sec:uvmaps} and discuss factors that affect the synthetic maps, including a discussion of uncertainties in the modeling. In Section~\ref{sec:results}, we examine results from this analysis, including the fraction of obscured star formation and comparison of \sfr{} calibrations. We discuss the factors that affect our results in Section~\ref{sec:discussion}. We conclude in Section \ref{sec:conclusions}.

\section{Data}
\label{sec:data}

\subsection
{\textit{GALEX} UV Images}
\label{subsec:galex_data}
We took UV observational data from the \galex{} Deep Imaging Survey \citep[DIS;][]{Martin2005a}, using the five tiles that cover the PHAT survey in the \fuv{} and \nuv{}. These tiles are listed in Table \ref{tab:galex_tileinfo}. We masked the tile edges and converted the count rate units, $\mathrm{cps}$, into flux according to

\begin{equation}
	f = U \left( \frac{\mathrm{cps}} {\mathrm{counts} \ssec^{-1} \,\mathrm{pixel}^{-1}} \right) \, ,
\label{eq:cpstoflux}
\end{equation}

\noindent where $U$, the \galex{} unit response, is given in Table~\ref{tab:galex_filters}. A small amount of background UV flux was present in the \fuv{} and \nuv{} mosaics, primarily due to scattering of UV photons from hot foreground stars in the Galaxy. We estimated the background as follows:
In each of the five tiles, we measured the mean flux in four $0.06 \times 0.06$ square degree apertures in off-galaxy areas relatively devoid of stars. The centers of each of the twenty apertures are listed in Table \ref{tab:galex_tileinfo} next to the appropriate tile. We then averaged these twenty measurements to get our estimate of the background flux and subtracted this value from all pixels in the corresponding mosaic. The subtracted background in the \fuv{} and \nuv{} images was $3.054\pm0.601\times10^{-16}$\uflambda{} and $2.386\pm0.678\times10^{-16}$\uflambda{}, respectively.

%-- TABLE --%
\begin{deluxetable*}{cccccl}
\tabletypesize{\footnotesize}
\tablecaption{\textit{GALEX} Observations.}
\tablewidth{0pt}
\tablehead{
    \colhead{Tilename} &
    \colhead{RA} & 
    \colhead{dec} &
    \multicolumn{2}{c}{exposure time (s)} &
    \colhead{Background Aperture Centers} \\
    \colhead{} &
    \colhead{} &
    \colhead{} &
    \colhead{FUV} &
    \colhead{NUV} &
    \colhead{$^\circ$ (J2000)}
}
\startdata
\texttt{PS\_M31\_MOS00} & 10.675 & 41.267 & 9863.55 & 95842.35 & (11.02, 40.78), (11.25, 41.02), (10.47, 41.80), (9.94, 41.39) \\
\texttt{PS\_M31\_MOS07} & 10.600 & 42.350 &  7418.25 & 50119.15 & (10.06, 42.13), (10.82, 42.21), (10.82, 42.73), (9.92, 42.48) \\
\texttt{PS\_M31\_MOS08} & 11.350 & 42.200 & 7497.6 & 55819.3 & (11.77, 42.62), (12.02, 42.03), (10.70, 42.16), (10.93, 42.63) \\
\texttt{PS\_M31\_MOS09} & 12.170 & 42.032 & 5431.7 & 48263.9 & (11.85, 41.98), (12.38, 42.21), (11.98, 41.49), (12.48, 42.53) \\
\texttt{PS\_M31\_MOS10} & 11.220 & 41.370 & 6561.5 & 51551.05 & (11.83, 41.40), (11.76, 41.64), (11.48, 40.99), (11.07, 41.04)
\enddata
\tablecomments{Tilename, RA, dec, exposure time in the \fuv{} and \nuv{}, and the centers of the four apertures in which the background was measured for each tile used to create the \galex{} UV mosaics.}
\label{tab:galex_tileinfo}
\end{deluxetable*}

\subsection{PHAT SFHs}
\label{subsec:sfhs}
To model UV flux, we use the PHAT spatially-resolved recent \sfh{s} determined in \citet{Lewis2015a}. We briefly describe their derivation here, but refer readers to the original paper for details. Each brick in the PHAT survey was divided into 450 regions on a uniform $15 \times 30$ grid with a total of \about9000 regions across the survey area, excluding the bricks closest to the crowded bulge area (see Figure \ref{fig:map}). Each region was approximately $24\arcsec \times 27\arcsec$ (100 pc $\times$ 100 pc, projected; 100 pc $\times$ 400 pc, deprojected). For each region, they modeled the optical (F475W and F814W) CMD using the fitting code \texttt{MATCH} \citep{Dolphin2002a}, which compares the observed CMD with many synthetic CMDs for composite stellar populations over a range of ages and metallicities. As described in detail in \citet{Lewis2015a}, the \sfh{s} were derived with the following assumptions:

\begin{enumerate}
\item A \citet{Kroupa2001a} IMF.
\item Padova isochrones \citep{Marigo2008a} with updated asymptotic giant branch (AGB) tracks \citep{Girardi2010a}.
\item A distance modulus of 24.47, corresponding to a distance of 783 kpc \citep{McConnachie2005a}.
\item A binary fraction of 0.35 with a uniform mass ratio distribution between 0 and 1.
\item Age resolution of 0.1 dex for $\log$(time) = 6.6 -- 9.9 and 0.25 dex for $\log$(time) = 9.9 -- 10.15 (In Section \ref{subsec:flux_modeling}, we correct the youngest time bin so that it extends to the present day.)
\item Metallicity resolution of 0.1 dex over the range $-2.3 \le \met{} \le 0.1$, with the requirement that \met{} increases with time.
\item A two-parameter extinction model consisting of a foreground component, $A_V$, applied evenly to all stars, and an additional differential component, $dA_V$ applied to stars following a uniform distribution, such that all stars in a region are extincted by some amount between $A_V$ and $A_V + dA_V$, optimized for each region individually \citep[see also][]{Simones2014a}. These values are used in a \citet{Cardelli1989a} extinction model with $R_V=3.1$.
\end{enumerate}

Additionally, the portion of each CMD with $\mathrm{F475W} > 21$ and $\mathrm{F475W} - \mathrm{F814W} > 1.25$ (red giant branch and red clump stars) was excluded from the fit \citep[see the CMDs in Figures 2 and 3 of][]{Lewis2015a}. This choice mitigates extinction effects from older stellar populations which are not well fit with the single step function described in item 7 above. As a result, the optimized extinction parameters correspond only to the dust associated with young, UV-emitting stars on the main sequence. The \sfh{s} are therefore limited to the last 500 Myr. We refer the reader to \citet{Dalcanton2015a} for a robust analysis of the dust in M31. The full \sfh{} analysis using the red features of the CMD and the \citet{Dalcanton2015a} dust analysis is the subject of an upcoming paper (B. Williams et al., in prep.).

We also note that the \sfh{s} of \citet{Lewis2015a} do not differentiate between cluster and field stars. As a result, UV light from young clusters is included in this analysis. However, stellar clusters account for only 4 -- 8\% of the star formation in M31 \citep{Johnson2016a}, and they therefore contribute only a few percent of the integrated UV light.

\section{Map Creation}
\label{sec:map_creation}

\subsection{Modeling Ultraviolet Flux}
\label{subsec:flux_modeling}

We used the \sfh{s} described in Section \ref{subsec:sfhs} to create spatially-resolved broadband \fuv{} and \nuv{} flux maps for the PHAT survey area. We modeled the flux in each region using a technique similar to that described in \citet{Johnson2013a}. We generated a set of simple stellar population (SSP) models and then weighted those models by the SFH to calculate the integrated SED for a given region. We ultimately determined the \fuv{} and \nuv{} flux and magnitude of each region from its modeled SED.

We first constructed a set of SSPs using the Flexible Stellar Population Synthesis (FSPS) code \citep{Conroy2009b, Conroy2010b}, assuming a \citet{Kroupa2001a} IMF and the Padova isochrones \citep{Marigo2008a} with updated AGB tracks \citep{Girardi2010a} and using the BaSeL 3.1 semi-empirical stellar SED library \citep{Westera2002a}. In FSPS, the Geneva tracks \citep{Meynet1994a} are used at $\log$(time)$<$6.6 for high-mass stars, supplemented with the Padova models for low mass stars. These choices are consistent with the \sfh{} determination. We constructed the SSPs with an age resolution of 0.025 dex over the range $\log$(age) from 5.500 to 10.175. We set the SSP metallicity to the mean metallicity over the last 100 Myr, as derived from the \sfh{}. If there was no star formation over that time range, we set the metallicity to that of the most recent time when the \sfr{} was non-zero.

To link the SSPs to the \sfh{}, we needed to reprocess the \sfh{}. The \texttt{MATCH} implementation of the Padova isochrones only reaches to $\log$(time) = 6.6, so we renormalized the SFR in the youngest age bin to reach time = 0, conserving the mass created in that time bin: 
\begin{equation}
\begin{split}
	\sfr(t=0 - 10^{6.7}\,\textrm{Myr}) =  \\
	\sfr(t=10^{6.6}-10^{6.7}&\,\mathrm{Myr})\times \left(1.0 - \dfrac{10^{6.6}}{10^{6.7}}\right).
\end{split}
\end{equation} 
We also increased the age resolution of the \sfh{} to \about6.5$\times 10^4$ yr. We determined this value by splitting the smallest time bin into 20 separate bins. In $\log$(time) space we then interpolated the SSP SEDs to the \sfh{} time points, weighted each interpolated SED by the SFR at each \sfh{} time point, and summed the SEDs to create the integrated intrinsic (i.e., dust-free) model SED. We note that we used the full 14 Gyr \sfh{} in this process despite the fact that \citet{Lewis2015a} stress that the \sfh{s} are only robust to \about500 Myr ago. We will discuss the effects of the \sfh{} timescale in the Appendix.

To include the effects of dust, we created the integrated, attenuated model SED in the same manner as the intrinsic SED, except that we reddened each individual SSP SED component before weighting by the mass. To redden the SED, we split each SSP SED into 30 identical component SEDs. Each component was then attenuated according to the \citet{Cardelli1989a} extinction curve with $R_V=3.1$, assuming a uniform random $A_V$ distribution drawn between \avsfh{} and \avsfh{} + \davsfh{}, where \avsfh{} and \davsfh{} are the best-fit parameters of the two-component extinction model used to derive the \sfh{} from the CMD \citep{Lewis2015a}. This was done to mimic the \sfh{} derivation, as described briefly in bullet point 7 of Section \ref{subsec:sfhs}.

We summed all of the attenuated components to create each region's integrated attenuated model SED. We note that the \citet{Cardelli1989a} extinction curve, which predicts the amount of extinction relative to that in the $V$ band as a function of wavelength, is based on the average \rv{} = 3.1 extinction curve for the Galaxy. Previous studies have shown that this extinction curve is applicable to M31 as a whole in both the UV \citep{Bianchi1996a} and the optical \citep{Barmby2000a} regimes. However, individual sight-lines may differ from \rv{} = 3.1 due to metallicity, gas-to-dust ratio, or star formation activity \citep{Clayton2015a}.

To determine the UV magnitude, we projected each model SED (both intrinsic and attenuated) onto the response curves for the \galex{} \fuv{} and \nuv{} filters to obtain absolute synthetic \fuv{} and \nuv{} magnitudes in the AB system. We converted the resulting absolute magnitude to apparent magnitude assuming a distance modulus of 24.47 \citep{McConnachie2005a}. 

Finally, we convert the magnitude to flux using:
\begin{equation}
	m = -2.5 \logten \left( \frac{f}{U} \right) + Z \; ,
\label{eq:magtoflux}
\end{equation}

\noindent where $U$ is the \galex unit response, and $Z$ is the zeropoint, given for each filter in Table \ref{tab:galex_filters}.

Throughout the text, we refer to the synthetic intrinsic (un-reddened) flux as \fxsfhz{} and the synthetic reddened flux as \fxsfh{}. The same nomenclature is used when referring to synthetic magnitudes as well.

We note that this process can be applied when modeling the flux at any wavelength, although appropriate care must be taken in the IR, where dust geometry and radiative transfer effects also need to be considered.

%-- TABLE --%
\begin{deluxetable}{lcc}
\tabletypesize{\footnotesize}
\tablecaption{\textit{GALEX} filter properties.}
\tablewidth{0pt}
\tablehead{
    \colhead{} &
    \colhead{FUV} &
    \colhead{NUV}
}
\startdata
Unit response, $U$ $(\times 10^{-15}\uflambda)$\tablenotemark{a} & 1.40 & 0.206 \\
\\
AB magnitude zeropoint, $Z$ & 18.82 & 20.08
\enddata
\tablenotetext{a}{\url{http://asd.gsfc.nasa.gov/archive/galex/FAQ/counts_background.html}}
\label{tab:galex_filters}
\end{deluxetable}

%% NEW SUBSECTION %%
\subsection{Creating Maps of Synthetic Flux}
\label{subsec:synmap_creation}

We used the modeled \fuv{} and \nuv{} fluxes in each region to create a single map of the PHAT survey area at each wavelength. We created these maps with Montage\footnote{
\url{http://montage.ipac.caltech.edu}
}, 
which aids in the combination of many FITS images into a single mosaic. Montage is flux-conserving and maintains the photometric and spatial fidelity of the input images.

Because the \sfh{s} were derived on a 15$\times$30 grid in each brick, we assembled the modeled flux values for each UV filter into a 15$\times$30 array to create a brick image. We then tied each brick image independently to a world coordinate system using the RA and Dec coordinates of the \sfh{} brick grid. We then let Montage determine the single template header that best describes the combination of input images and used that header to re-project each brick image to the same WCS. We co-added the re-projected images into a single FITS file to create full PHAT survey area maps of synthetic \fuv{} and \nuv{} flux. The pixels in the resulting co-added image have a scale of $23.7$\arcsec. Due to the reprojection, the input regions do not match one-to-one to the output image pixels. We do not background-match the synthetic images.

The rotation of the outer bricks with respect to the inner bricks results in a number of locations where pixels in one brick overlap with pixels in another brick, most notably in Bricks 9 and 11. We note, however, that three of the fields in B11 were not observed. These three fields overlap completely with fields from B09, decreasing the total amount of overlap. We compared the output reddened NUV flux values in 10 of these overlapping pixels. We found the agreement to be within 10 -- 15\% on average.

Assuming a distance modulus of 24.47 \citep{McConnachie2005a}, a disk inclination of $77\degree$ \citep[e.g.,][]{Roberts1966a, Brinks1984b, Walterbos1988a}, and a major axis position angle of $35\degree$ \citep{de-Vaucouleurs1995a}, the pixel scale deprojects to a linear size of 440\pc{}$\times$100\pc{} along the minor and major axes of M31, respectively (see orientation of scale bar in Figure \ref{fig:map}). The synthetic flux maps therefore have a resolution that is firmly in the sub-kpc regime. 

Montage creates a pixel weight map during the mosaicking process. All pixels are weighted according to the fraction of the pixel that is inside the PHAT footprint (i.e., pixels entirely inside the footprint have a weighting of 1, pixels entirely outside of the PHAT footprint have a weighting of 0, and pixels that are partially within the footprint have a weighting somewhere between 0 and 1). To remove possible border effects, we only analyze pixels that have a weight $w > 0.95$.

%% NEW SUBSECTION %%
\subsection{Turning \galex{} Observations into Maps}
\label{subsec:galex_map_making}
We also constructed maps of observed \galex{} flux (\fxobs), matched to the sampling of the modeled synthetic flux maps. The process was similar to that used for the synthetic maps. Starting with the background-subtracted images described in Section \ref{subsec:galex_data}, we reprojected the flux tiles to the same template header as the synthetic flux maps. We then background-matched the images by determining the differences in the images at their overlaps and fitting a plane to these difference images to model the background. Finally, we co-added the matched images to create the final observed maps.

%-- FIGURE --%
\begin{figure*}
\centering
\includegraphics[width=\textwidth]{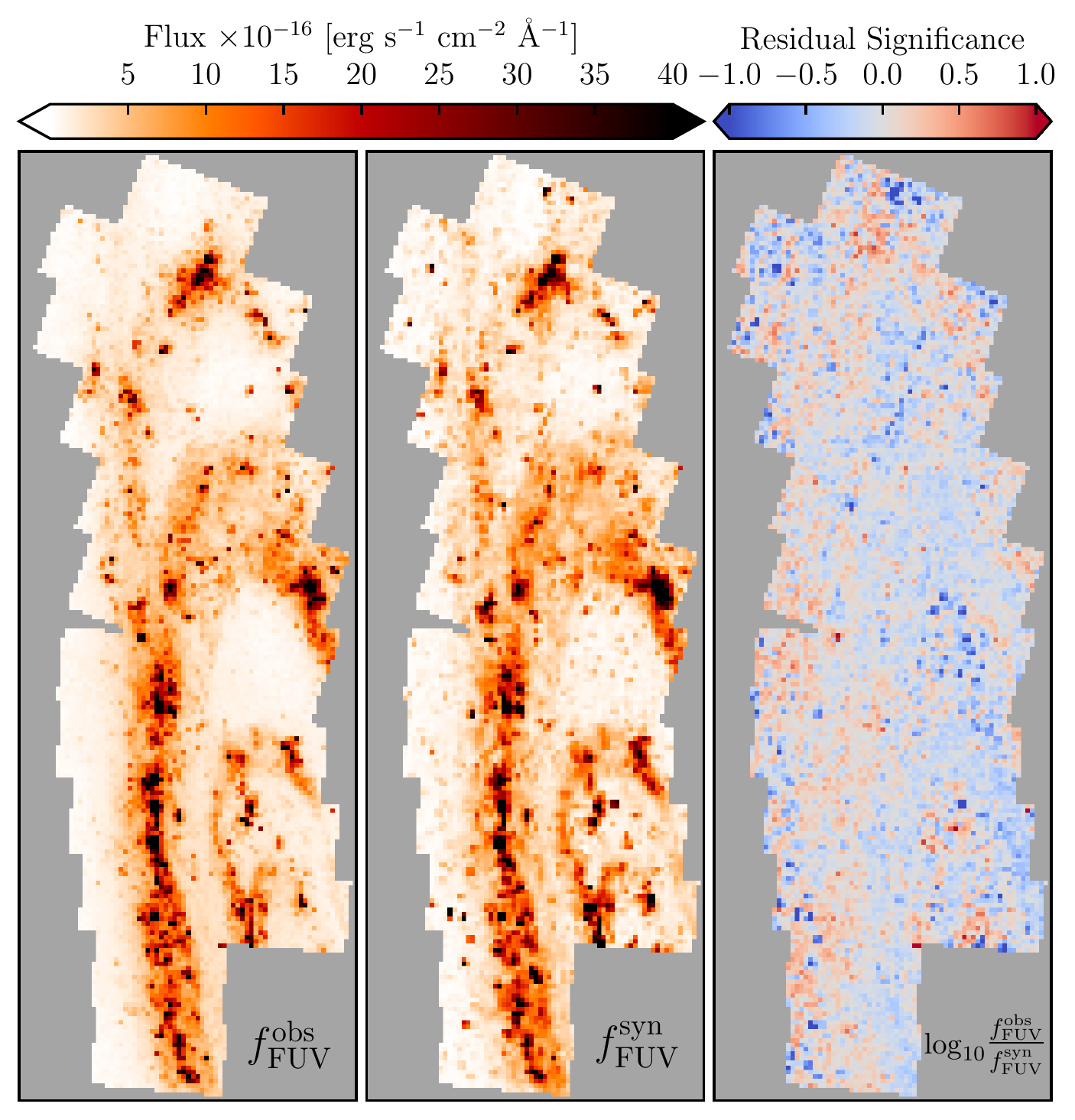}
\caption[Observed and modeled \fuv{} flux maps.]{The observed \galex{} \fuv{} flux, \ffuvobs{}, is shown in the left panel. The middle panel shows the synthetic flux derived from optical \sfh{s} and attenuated according to a Cardelli extinction model, \ffuvsfh{}.  These maps are on the same flux scale as indicated by the color bar above the left two panels. The right panel shows the significance of the residuals, plotted as $\log$(\fxobs{} / \fxsfh{}). Bluer colors are where the modeled flux is over-predicted and redder colors are where it is under-predicted.}
\label{fig:fuvmaps}
\end{figure*}

%-- FIGURE --%
\begin{figure*}[]
\centering
\includegraphics[width=\textwidth]{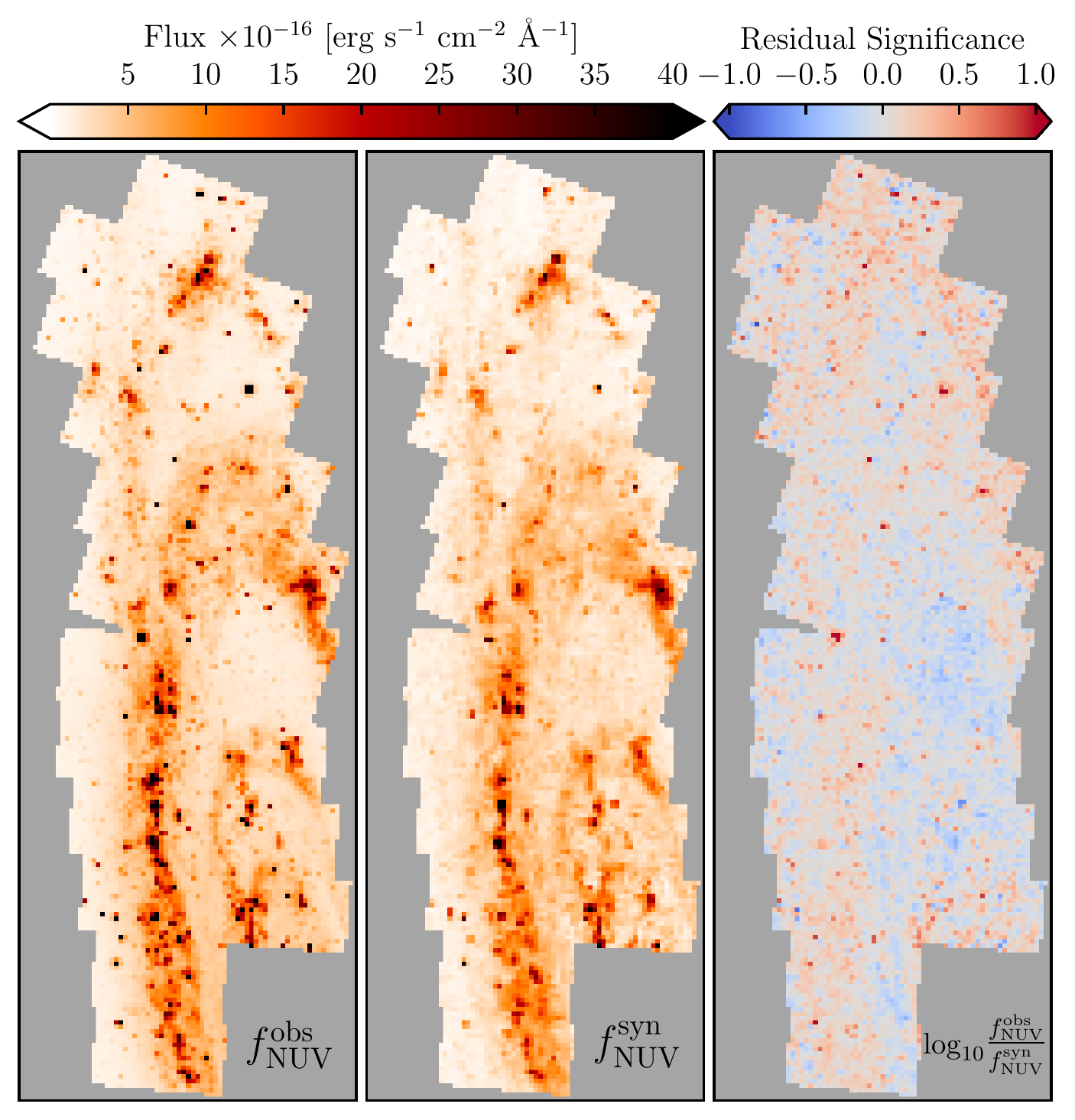}
\caption[Observed and modeled \nuv{} flux maps.]{Same as Figure \ref{fig:fuvmaps} except showing the \nuv{} results. The left panel shows the observed flux, \fnuvobs{}, the middle panel shows the modeled, attenuated flux, \fnuvsfh{}, and the right panel shows the significance of the residuals between the two.}
\label{fig:nuvmaps}
\end{figure*}

%% NEW SECTION %%
\section{Verification of the Optically-derived Synthetic Ultraviolet Maps of the PHAT Survey}
\label{sec:uvmaps}

\subsection{Presentation of the Maps}
\label{subsec:presentmaps}

In Figures \ref{fig:fuvmaps} and \ref{fig:nuvmaps} we show maps of observed \galex{} flux (\fxobs{}; \S\ref{subsec:galex_map_making}), synthetic attenuated flux (\fxsfh{}; \S\ref{subsec:flux_modeling} and \S\ref{subsec:synmap_creation}), and the significance of the difference between the two over the PHAT survey area for the \fuv{} and \nuv{}, respectively. All figures have the same stretch.

Figures \ref{fig:fuvmaps} and \ref{fig:nuvmaps} show remarkable qualitative agreement between the synthetic and the observed fluxes, indicating that the synthetic attenuated fluxes derived from the optical CMDs do an excellent job at reproducing the observed fluxes. All of the main features of the \galex{} maps are reproduced in the synthetic maps, including the 10 kpc ring, the ring features at 5 and 15 kpc, and the individual star-forming regions within the ring, as well as the OB associations in Bricks 15 and 21 (the bright features located at the top and right side of the left two panels in Figures \ref{fig:fuvmaps} and \ref{fig:nuvmaps}; also see Figure \ref{fig:map} for brick numbering). The agreement is especially good in the \fuv{} map, but also apparent in the \nuv{} map, despite the fact that some features are not as defined in the synthetic map. We emphasize that we have used \emph{optical} colors and magnitudes to derive the \emph{ultraviolet} fluxes. 

There are, however, distinct differences between the \fxsfh{} and \fxobs{} maps. For example, the \fxobs{} maps show some point-like sources that do not appear in the \fxsfh{} maps. Inspection of the original images shows that these sources are largely Milky Way foreground stars. Foreground stars are typically at a very different absolute magnitude than M31 stars, and it is unlikely that any foreground stars were inadvertently included in the CMD modeling. Additionally, foreground stars don't produce features in the synthetic maps because the fluxes in those maps are derived from the \sfr{} of a \emph{distribution} of stars, rather than single stars, so while a single bright star may dominate the pixel in the observed map, it has less of an effect in the modeled map because it is averaged out in the total \sfh{} from which the modeled flux is derived. 

There are also a number of bright pixels in the modeled maps that are not in the observed maps. These mostly correspond to photometric artifacts (primarily diffraction spikes) that remained in the PHAT photometry\footnote{
We note that the \sfh{s} derived in \citet{Lewis2015a} used the first version of the PHAT photometry. The photometry has since changed, using different quality cuts and resulting in overall improvement. Given the large computing expense to run the \sfh{s} on the entire grid and the minimal difference it would make for the \sfh{s}, it was not deemed worthwhile to re-run the \sfh{s} on the most recent photometry.
} 
after application of the quality cuts \citep{Dalcanton2012a}. These artifacts increase the apparent stellar mass measurement in the \sfh{}, which increases the synthetic flux. This contamination affects $<$5\% of the pixels.

\begin{figure}[t]
\centering
\includegraphics[width=0.95\columnwidth]{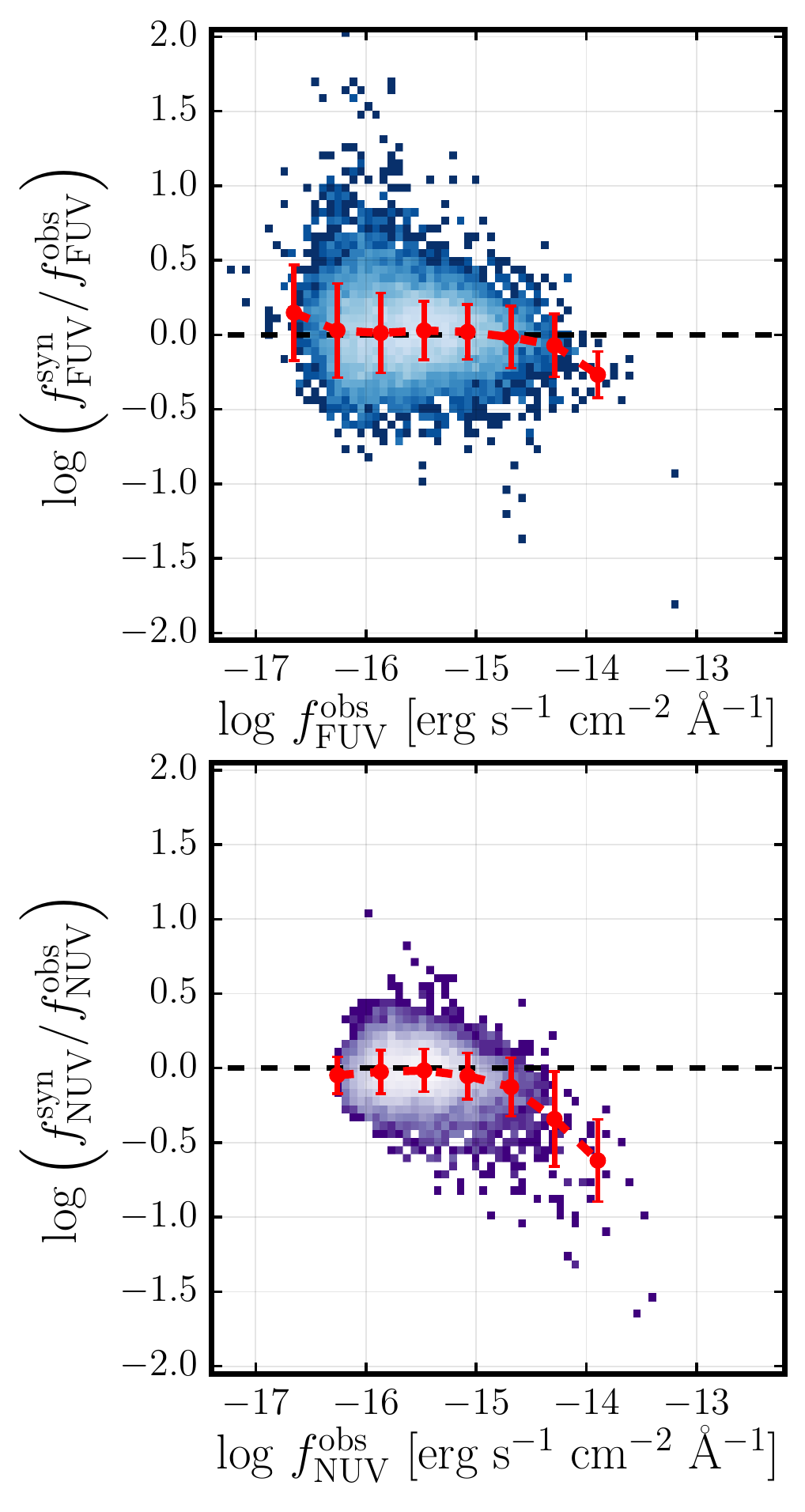}
\caption[Ratio of observed and modeled flux]{The log ratio of the modeled flux to the observed flux as a function of observed flux for all pixels that fall inside the PHAT footprint. The red circles in each panel show the running median with the standard deviation given by the error bars on each point. The median log ratios are 0.005 and -0.027 in the \fuv{} and \nuv{}, respectively. The standard deviation is 0.24 in the \fuv{} and 0.16 in the \nuv{}.}
\label{fig:fluxratiocompare}
\end{figure}

%-- FIGURE --%
\begin{figure*}[]
\centering
\includegraphics[width=\textwidth]{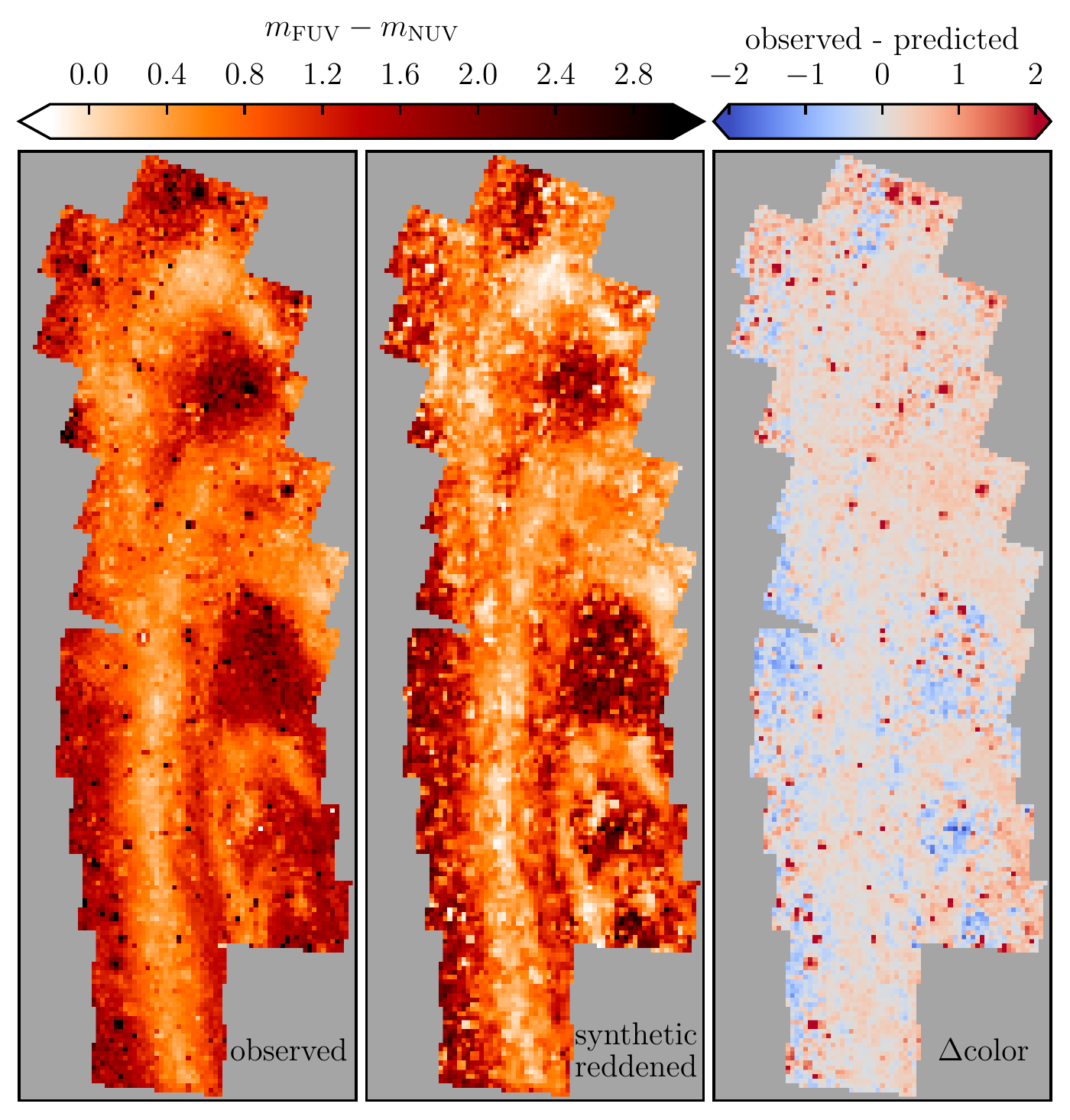}
\caption[UV color maps.]{Maps of UV color. The left panel shows the observed color,  \fuvnuvobs{}, the middle panel shows the modeled, attenuated color, \fuvnuvsfh{}, and the right panel shows the difference between the two}
\label{fig:uvcolormaps}
\end{figure*}

In Figure \ref{fig:fluxratiocompare}, we present a quantitative comparison of the synthetic and observed fluxes shown in Figures~\ref{fig:fuvmaps} and \ref{fig:nuvmaps}. We plot the ratio of the modeled reddened flux to the \galex{} flux (\fxsfh{}/\fxobs{}) as a function of \galex{} flux, including the running median and standard deviation. The overall median and standard deviation are $\log_{10}(\ffuvsfh{}/\ffuvobs{}) = 0.03 \pm 0.24$ and $\log_{10}(\fnuvsfh{}/\fnuvobs{}) = -0.03 \pm 0.16$. Together with the qualitative comparison in Figures~\ref{fig:fuvmaps} and \ref{fig:nuvmaps}, these numbers indicate that in both filters, \fxsfh{} is consistent with \fxobs{}. The largest average discrepancy occurs at high observed flux, where the number of regions is very small, and are likely due to foreground stars or other photometric artifacts. 

The \fuv{} relation is flat across the range of observed fluxes. The scatter on the best-fit values decreases as the observed flux increases, ranging from 0.3 dex at the low end to 0.1 dex at the high end.  The two regions at the highest \fuv{} flux actually correspond to a single bright star located at the intersection of B12 and B14, clearly visible in Figure 1 and in the left panel of Figure 2. On a region-by-region basis, regions of low observed flux show the largest deviations between the observed and synthetic flux.

In contrast, the \nuv{} data are much tighter, showing a scatter of \about0.1 dex at low observed fluxes and increasing to \about0.2 dex at the brighter end. There is also a slight trend in the \nuv{} data. The synthetic flux is generally in very good agreement in regions of low observed flux but under-estimated in regions of high observed flux. This behavior can be seen in the right panel of Figure \ref{fig:nuvmaps} where the regions between and outside of the ring features (low observed flux) are gray to light blue in color, indicating only slightly more synthetic flux than observed flux, and the ring regions (high observed flux) are red, indicating more observed flux than synthetic flux. 

At the high flux end of the \nuv{} distribution, the inconsistency must lie with the modeled data. One possibility could be our treatment of dust, because the regions of high flux coincide with the dustiest regions. However, both the \fuv{} and \nuv{} are treated the same way and this problem does not affect the \fuv{}, which should be even more sensitive to errors in the dust model; we cannot assume that the dust is too high in the \nuv{} and just right in the \fuv{}. We note, however, that the \nuv{} band includes the 2175\AA{} bump. If we were to reduce the strength of the bump at the same $R_V$ \citep[e.g.,][]{Conroy2010d}, this change would increase the flux in the \nuv{}, bringing the ratio of synthetic to observed flux closer to one but leaving the \fuv{} flux unchanged. A variation in bump strength would also have little effect on the low-flux regions which are also generally low-dust regions and are therefore less affected by the robustness of the extinction curve. 

A final factor that could contribute to the trend seen in the \nuv{} data is timescale. When we constructed the set of SSPs, we set the metallicity of each SSP to the mean metallicity of the \sfh{} over the last 100 Myr. The 100 Myr timescale is based on the average lifetime of the O and B stars that emit in the \fuv{}, whereas the timescale for \nuv{} emission is longer at 300 Myr. We tested using 300 Myr as the baseline over which to determine the mean metallicity for each SSP and found no difference in the results. M31's metallicity has not changed significantly over the last few hundred Myr. Therefore, metallicity is not the reason for the observed trend.

\begin{deluxetable*}{ccccccccc}
\tabletypesize{\footnotesize}
\tablecaption{Fraction of pixels within 1, 2, or 3 Sigma of One-to-One}
\tablecolumns{9}
\tablewidth{0pt}
\tablehead{
    \colhead{Bin Center} &
    \multicolumn{2}{c}{N} &
    \multicolumn{2}{c}{$1\sigma$} &
    \multicolumn{2}{c}{$2\sigma$} & 
    \multicolumn{2}{c}{$3\sigma$} \\ 
    \colhead{$\log$ $\left(\uflambda{}\right)$} & 
    \colhead{FUV} & \colhead{NUV} & 
    \colhead{FUV} & \colhead{NUV} & 
    \colhead{FUV} & \colhead{NUV} & 
    \colhead{FUV} & \colhead{NUV}
}
\startdata
-16.65 &  123    & 0       & 0.70 & NA   & 0.93 & NA   & 0.98   & NA \\
-16.25 & 1518   & 288   & 0.78 & 0.66 & 0.95 & 0.94 & 0.98 & 0.99 \\
-15.87 & 2785   & 3192 & 0.75 & 0.69 & 0.95 & 0.96 & 0.99 & 1.0 \\
-15.47 & 3132   & 4404 & 0.73 & 0.72 & 0.95 & 0.95 & 0.99 & 0.99 \\
-15.08 & 1788   & 1780 & 0.72 & 0.73 & 0.95 & 0.92 & 0.99 & 0.98 \\
-14.68 & 632     & 376   & 0.78 & 0.65 & 0.95 & 0.89 & 0.98 & 0.97 \\
-14.29 & 122     & 66     & 0.70 & 0.48 & 0.92 & 0.82 & 0.99 & 0.92  \\
-13.99 & 12       & 14     & 0.33 & 0.00 & 0.67 & 0.50 & 0.92 & 0.79
\enddata
\tablecomments{The first column shows the center of the bin of width 0.4 dex. Columns 2 and 3 indicate the number of pixels that fall within that bin. Columns 4 and 5 show the fraction of pixels in that bin that lie within 1$\sigma$ of the one-to-one line (the dashed line in Figure \ref{fig:fluxratiocompare}). Columns 6 and 7 show the fraction of pixels that lie within 2$\sigma$ of the one-to-one line. Columns 8 and 9 show the fraction of pixels that lie within 3$\sigma$ of the one-to-one line.  }
\label{tab:stats}
\end{deluxetable*}

We convert flux to magnitude and compare maps of observed and synthetic UV color (\fuvnuv{}) in Figure \ref{fig:uvcolormaps}. The left panel shows the observed color, the middle panel shows the predicted, reddened color, and the right panel shows the difference between the two. The main difference between the left and middle panels is that the predicted map is bluer in the star-forming regions along the 10-kpc ring. This is because the synthetic \nuv{} flux is slightly under-predicted in the high flux, star-forming regions, resulting in fainter \nuv{} magnitudes and therefore bluer colors. This is confirmed in the third panel, which shows the difference between the observed and predicted colors.

Despite the systematic trend in the \nuv{} and the larger scatter in the \fuv{}, we find that 70--75\% of pixels in both the \fuv{} and the \nuv{} fall within $1\sigma$ of the one-to-one line and $>$97\% fall within $3\sigma$. These percentages do not include the brightest single bin in the \fuv{} and the brightest two bins in the \nuv{} where those numbers are significantly lower. The number of points in each bin as well as the percentage that fall within 1, 2, or 3$\sigma$ are given in Table \ref{tab:stats}. While we do not match all points perfectly, the degree of agreement is still impressive.

The overall agreement between \fxsfh{} and \fxobs{} confirms that our modeling procedure is generally robust and justifies the assumptions we used to model the flux, including the assumed IMF, stellar models and spectral library, and the extinction model. While the modeling can certainly be made more complex, it is reassuring to know that we can use all of this knowledge to derive \sfh{s}, synthesize SEDs, and successfully recreate detailed maps in the UV, all from photometry in only two optical bands.

\subsection{Issues Affecting the Synthetic Ultraviolet Maps}
\label{subsec:issues}

The above comparisons strongly support the overall reliability of the \sfh{s} presented in \citet{Lewis2015a}, particularly the dust parameters derived in the \sfh{} fitting process. Therefore, before we proceed to a discussion of the un-reddened maps, we highlight some additional considerations that could affect the modeled flux. 

The \galex{} data primarily suffers from Poisson uncertainties due to the exposure time. For each of the five DIS images in this study, we  assume an average exposure time of $7\times 10^3$ s in the \fuv{} channel and $6\times 10^4$ s in the \nuv{} channel. The uncertainties are only a few percent at $f^{obs}$\about $10^{-16} \uflambda$. The \nuv{} uncertainties are an order of magnitude smaller. 

The scatter in Figure~\ref{fig:fluxratiocompare} is much larger than the above Poisson uncertainties allow. As a result, the scatter seen in Figure~\ref{fig:fluxratiocompare} is most likely dominated by the modeling process. We now discuss possible sources of uncertainty in our synthetic data. 

There are four major factors that could contribute to the scatter observed in the synthetic fluxes: (1) \sfh{} uncertainties, (2) IMF sampling incompleteness, (3) differences between the dust model and the physical properties of the dust, and  (4) stellar evolution model uncertainties. Because each of these effects is coupled to the others (e.g., the \sfh{} of each region is dependent on the stellar models used and is also degenerate with the dust in that region), it is difficult to constrain the uncertainties from each source separately to high precision. Instead, we provide a qualitative discussion of each source and its effect on the resulting flux.

\subsubsection{SFH Uncertainties}

We first consider the role of uncertainties in the \sfh{}. At the native time resolution of 0.1 dex, the uncertainties on the SFR in a single time bin can range from $\gtrsim$10--100\%. The exact number depends on the number of luminous main sequence stars in each region \citep{Lewis2015a}, which will affect the certainty of star formation happening in one time bin as opposed to the adjacent time bin. These uncertainties naturally decrease when averaged over larger region sizes or longer timescales. 

The uncertainties on the \sfr{} in \citet{Lewis2015a} were determined using a hybrid Markov Chain Monte Carlo routine \citep{Duane1987a} to produce a sample of 10,000 \sfh{s}. To determine the magnitude of the uncertainties propagated through from the \sfh{s}, we ran 1000 of these \sfh{s} in each region through our SED modeling routine. We take the uncertainty on the flux to be the distribution of fluxes derived from this sample. We note that the uncertainties derived in this manner will be upper limits. Most of the stars that are responsible for the observed UV flux are present in the CMDs, and these stars also account for much of the observed star formation. This is especially true in the \fuv{}, where it is expected that younger populations, which we observe on the CMD, are responsible for most of the observed flux. 

\begin{figure*}[]
\centering
\includegraphics[width=\textwidth]{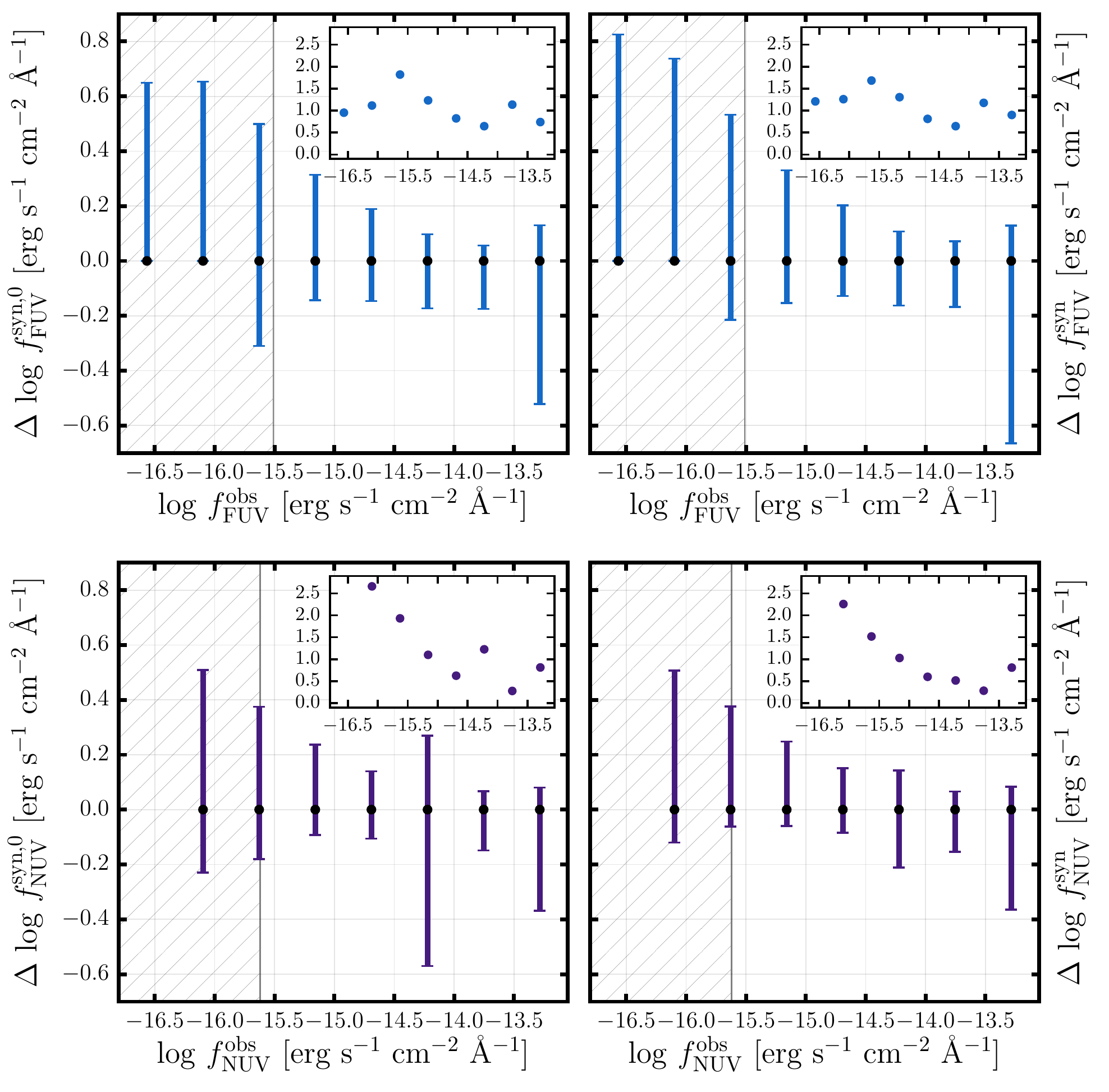}
\caption[Uncertainties on the modeled flux from the \sfh{s} as a function of observed flux]{Uncertainties from the SFH derivation on the dust-free and reddened synthetic flux as a function of observed flux. The top panels show the \fuv{} uncertainties and the bottom panels show the \nuv{} uncertainties. Dust-free flux is on the left and reddened flux is on the right. The gray vertical line indicates the total background level (Section \ref{subsec:galex_data}) and therefore the hatched region indicates the flux levels at which \galex{} noise takes over. The insets show the ratio of the uncertainties presented in this figure to the scatter shown in Figure \ref{fig:fluxratiocompare}.}
\label{fig:fluxuncs}
\end{figure*}

In Figure \ref{fig:fluxuncs}, we show the uncertainties in the modeled flux as a function of observed flux for both the reddened and dust-free synthetic flux. The uncertainties are highly asymmetric, especially at the low-flux end because of asymmetries in the \sfh{} uncertainties. When the best-fit \sfr{} in a time bin is zero, the uncertainty on that \sfr{} is only positive. The regions that populate the low-flux end of this figure tend to have little or no star formation in the most recent 100 Myr. Over the 100 Myr timescales of interest in this paper, the random uncertainties on the synthetic flux due to the \sfh{} derivation are up to an order of magnitude at the very faint end, though only a factor of a few or less at the bright end. The inset in each subpanel shows the ratio of the total uncertainties shown in Figure \ref{fig:fluxuncs} to the size of the scatter in Figure \ref{fig:fluxratiocompare}. 

\begin{figure*}[]
\centering
\includegraphics[width=\textwidth]{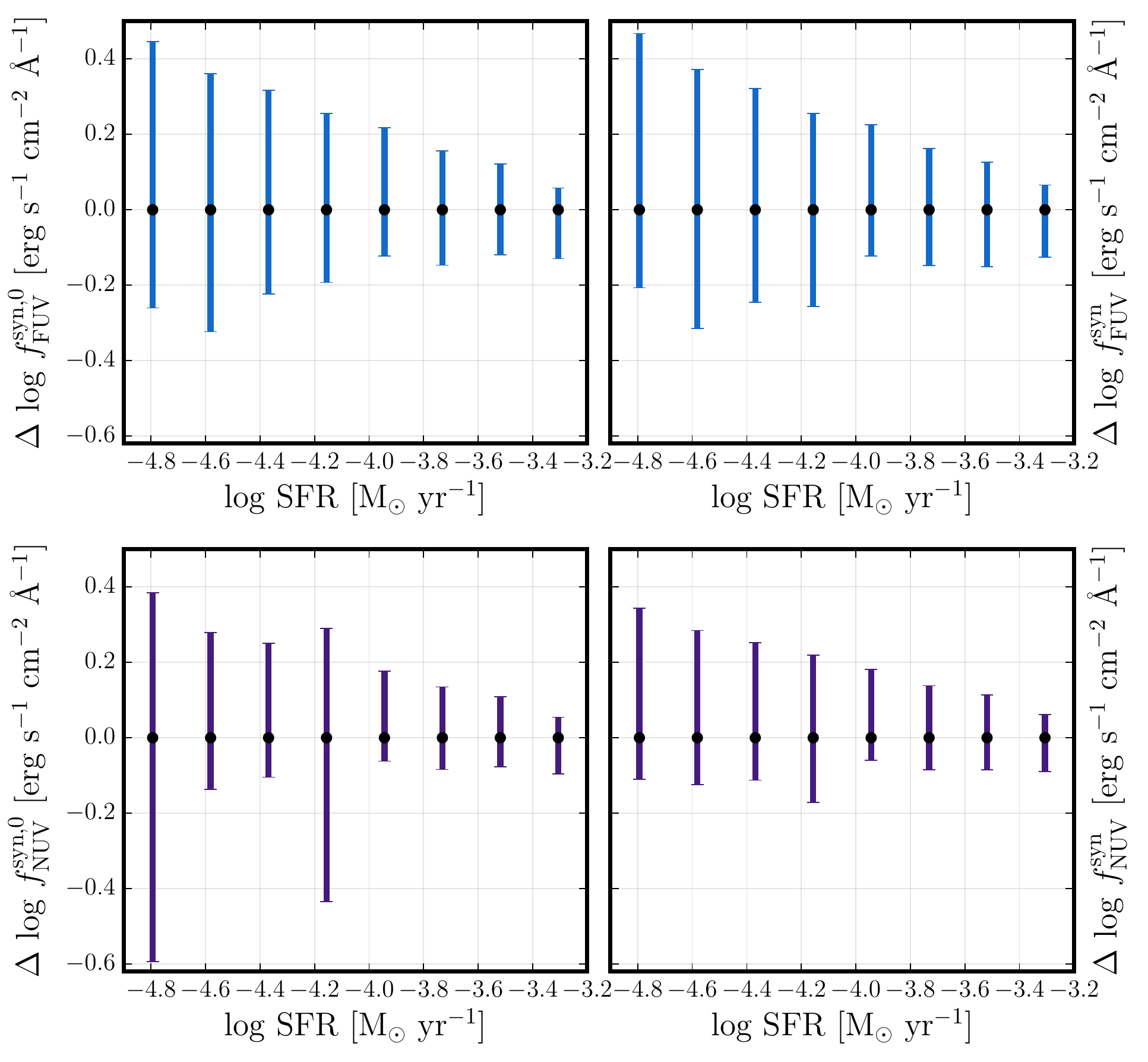}
\caption[Uncertainties on the modeled flux from the \sfh{s} as a function of \sfr{}]{Uncertainties from the SFH derivation on the dust-free and reddened synthetic flux as a function \sfr{} averaged over the last 100 Myr. The top panels show the \fuv{} uncertainties and the bottom panels show the \nuv{} uncertainties. Dust-free flux is on the left and reddened flux is on the right. We have only included \sfr{s} above $10^{-5}$ \sfrunit{}. Uncertainties at lower \sfr{s} (not on this figure) are 0.8 -- 1.0 dex in the \fuv{} and 0.4 -- 0.5 dex in the \nuv{}.}
\label{fig:fluxuncs_sfr_recent}
\end{figure*}

We repeat the same exercise in Figure \ref{fig:fluxuncs_sfr_recent}, except we plot the uncertainties as a function of \sfr{} averaged over the most recent 100 Myr. Similar to Figure \ref{fig:fluxuncs}, the uncertainties are asymmetric. As expected, they generally decrease toward higher \sfr{}.

\subsubsection{Dust Models}

We next consider uncertainties in the dust model used to model the \sfh{s} and the flux. As described in \citet{Lewis2015a}, we modeled dust in the optical CMD-derived \sfh{s} with a two-component top-hat model designed to account for constant extinction along the line of sight, and differential extinction internal to the galaxy. The CMD residuals suggest that this model is reasonable  for the young stars, but extrapolation of this model to the UV requires assumptions about the attenuation curve and the 2175 \AA{} bump. In the modeling process, we assume that the attenuation curve, $R_V$, and the strength of the 2175 \AA{} bump are constant across the disk. However, it is possible that these have spatial variation, and this could lead to scatter in the flux ratios. Robust SED fitters, such as the BEAST \citep{Gordon2016a} may be able to measure such region-to-region variation. 
Additionally, scattering of light by dust (Section \ref{subsec:scattered_light}) may alter the effective attenuation curve in a given region, producing extra flux that does not originate in that region and giving the attenuation curve a shape that is different from that of a perfectly known extinction curve.

\subsubsection{Incomplete Sampling of the IMF}

A third source of uncertainty is incomplete sampling of the stellar IMF. The UV light in a region is highly dependent on the number of massive stars. In modeling the \sfh{s}, we assumed a fully-sampled IMF. When the \sfr{} $\lesssim10^{-4}$ \sfrunit{}, however, the IMF may not be fully populated in that there are too few massive stars. This is most likely to occur in regions that are physically small or regions where the \sfr{} is low or non-existent. That said, low flux regions tend to contain mostly older stellar populations. As such, stochastic effects do not make as much of a difference because the UV is coming from stars further down the luminosity function. In high-flux and high-\sfr{} regions, a scale of 30 -- 40 pc (approximately the size of a 4000\msun{}, 5 Myr old cluster) appears to be large enough to avoid IMF sampling issues \citep{Boquien2015a}. Additionally, several studies have shown that stochastic sampling of the IMF -- even at very low SFR -- has less impact on the \fuv{} and \nuv{} fluxes because they result from integration over a wide mass range and are less dependent on the most massive stars than an ionizing luminosity such as H$\alpha$ \citep[e.g.,][]{Lee2009a, Lee2011a, da-Silva2012a, Johnson2013a, da-Silva2014a}. 

In general, the 100 pc $\times$ 400 pc regions in this study are more than large enough to avoid effects of stochastic sampling in high \sfr{} regions. In lower \sfr{} regions, stochastic sampling could potentially affect our results. However, incomplete sampling of the IMF does not overwhelmingly contribute to the observed scatter in the flux ratios. The \fuv{} and \nuv{} fluxes are integrated over a stellar mass that reaches down to $\sim$3 -- 5 \msun{} on the main sequence. Additionally, stochastic effects of the IMF are incorporated into the uncertainties on the \sfh{s} as described in \citet{Lewis2015a}. Consequently, it is unlikely that incomplete sampling of the IMF contributes significantly to the scatter in the flux ratios beyond the uncertainties already accounted for in the SFH measurements.

\subsubsection{Stellar Models}

The last source of uncertainty is from deficiencies in the stellar models used to derive the \sfh{s} and to model the flux. Just as the number of massive stars affects the observed UV light, discrepancies between the synthetic and true UV properties of these massive, metal rich stars could introduce scatter into the flux ratios. This uncertainty is the most challenging to quantify. The fidelity of the stellar models will be the same in all locations, but the impact of the models will vary from region to region due to variation with stellar type and mass and therefore with \sfh{} and IMF sampling. Perhaps most importantly, stellar models in the UV are poorly calibrated due to the paucity of massive, metal-rich stars in the local universe.

\subsubsection{Summary of Uncertainties}

Based on the above effects, the scatter in the flux ratios shown in Figure~\ref{fig:fluxratiocompare} is dominated by the uncertainties in the \sfh{}, which include scatter due to incomplete sampling of the IMF.

\subsection{Emission Timescales}
\label{subsec:emissiontimescales}

Fluxes in the \fuv{} and \nuv{} are often attributed to timescales of $<$100 Myr and $<$300 Myr, respectively, because most of the UV emission comes from the hottest, most massive stars that dominate younger stellar populations \citep{Kennicutt2012a}. It has also been shown that older populations emit in the UV as well \citep[e.g.,][]{OConnell1999a}, with often as much as 20-30\% of the total emission in the \fuv{} or \nuv{} coming from these stars \citep[e.g.][]{Johnson2013a}. The contribution from older populations is generally negligible in regions of high recent star formation. However, in regions of low SFR, this contribution can be non-negligible, particularly in the \nuv{}. Unfortunately, the CMD modeling in \citet{Lewis2015a} excluded regions of the CMD in the fit that are most sensitive to old stellar populations, and thus our ability to model the UV contribution from older star formation is limited. Including this contribution would be most likely to increase the low \sfr{} intensity region fluxes while leaving the high \sfr{} intensity regions unaffected.

\subsection{Scattered Light}
\label{subsec:scattered_light}

A portion of a galaxy's UV light can be diffuse and not associated with an obvious star-forming region. This diffuse UV emission is found between visible spiral arms or rings. In some cases, this fraction can be quite significant \citep[up to 65\% in M33;][]{Thilker2005a}. There are a number of suggestions for the origination of this diffuse light, including the dispersal of B--type stars after the dissolution of their natal clusters, a low level of diffuse star formation, or the scattering by dust of UV photons produced in bright star--forming regions \citep{Marcum2001a, Crocker2015a}.  

This scattered light may redistribute the flux in the observed UV maps in a manner that is not captured in our synthetic maps. We can, however, use our modeled data to examine the extent of scattered light in M31. The right panel in Figure~\ref{fig:fuvmaps} shows the log ratio of the observed to the modeled reddened \fuv{} flux. Blue pixels have more synthetic flux than observed with \galex{} and red pixels have more observed flux than synthetic flux. We have predicted slightly more synthetic flux than is observed (blue pixels) in just over half of the regions. However, the regions in which we have under-predicted the flux have a larger offset than the regions in which we have over-predicted the flux. This excess \galex{} flux occurs primarily along the eastern side of the 10 kpc ring as well as near two large OB associations in Bricks 15 \& 21. This discrepancy is particularly interesting because it is found near regions of very high, recent star formation. The many star-forming regions along the northeastern section of the ring are responsible for the bump in the star formation rate seen \about50 Myr ago \citep{Lewis2015a}. The excess observed flux seen near these highly star-forming regions could be an indication that some of this UV light has been scattered by dust into regions where no star formation is occurring. Further analysis of scattered light is beyond the scope of this paper.

\subsection{Comparison with PHAT F275W data}
\label{subsec:f275w_comparison}

The PHAT survey consists of data in six filters, two each in the NUV, the optical and the near-infrared. In the \sfh{} derivation \citep{Lewis2015a}, we used only the optical data as they are the deepest and provide the greatest leverage on the recent \sfh{}. In theory, the ultraviolet filters (F275W and F336W) would also provide important constraints on the recent \sfh{}. In practice, only the brightest main sequence stars have measurements in the UV filters, which greatly reduces the age range over which the \sfh{} can be derived. Therefore, the UV filters were not used in the \sfh{} derivation.

\begin{figure*}[]
\centering
\includegraphics[width=0.95\textwidth]{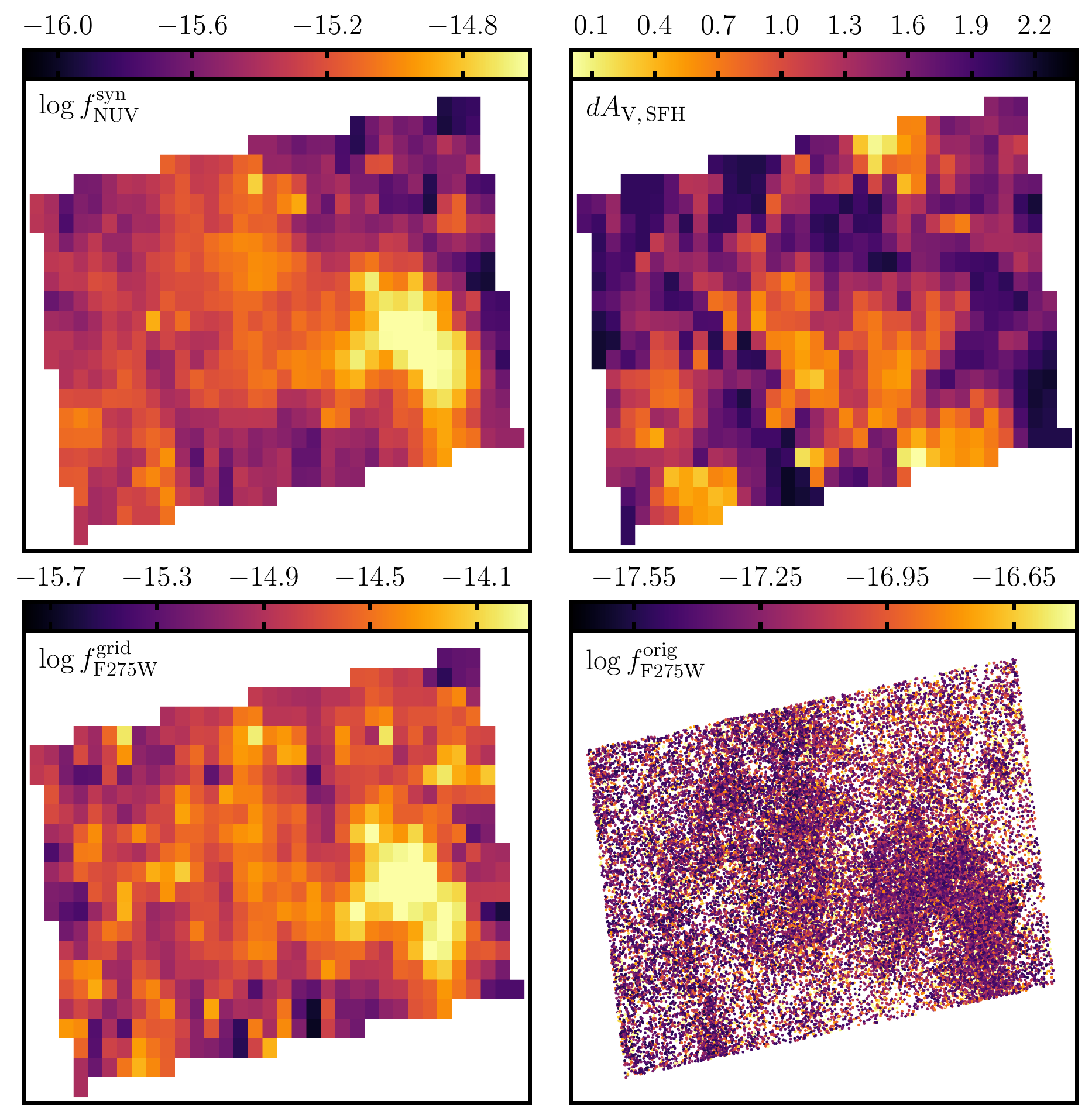}
\caption[Comparison with PHAT F275W data in B15.]{Comparison of synthesized NUV data with PHAT F275W \texttt{gst} data in B15. Synthetic, reddened NUV image (top, left), optimized internal reddening parameter determined from the \sfh{} derivation, \davsfh{} (top, right), PHAT F275W flux from individual stars binned to 100 pc (bottom, left), PHAT F275W flux, individual stars (bottom, right). }
\label{fig:nuv_f275w_compare}
\end{figure*}

We can, however, compare the PHAT UV data, in particular the F275W data, with the synthetic, reddened NUV maps derived from the optical data. In Figure \ref{fig:nuv_f275w_compare}, we show several maps of B15 including the synthetic, reddened \nuv{} flux (top, left), the internal reddening derived from the \sfh{} fitting process, \davsfh{} (top, right), and the F275W flux in individually detected stars, both gridded to the same scale as the UV maps (bottom, left) and at the individual star resolution (bottom, right). The synthetic \nuv{} and F275W flux images are colored according to the log of the flux in each band.

We chose B15 because it sits on the 10 kpc ring and therefore contains both active star formation and dust. The most obvious region of star formation is the bright region in the lower right of the brick \citep[OB 54;][]{vandenBergh1964a}. In Figure \ref{fig:nuv_f275w_compare}, both the synthetic \nuv{} and the gridded F275W flux in the left panels show a region of high flux (yellow in color) in that region. In the bottom right panel, the density of stars detected in F275W is much greater in that region and along the ring feature than in other parts of the brick.

The F275W data comes from the \texttt{gst} catalog, which contains only stars with reliable measurements (high signal-to-noise and low crowding) in each band. We chose to use the individually resolved stars rather than the full brick images of B15 for a number of reasons. There are only two exposures per UVIS band, making it difficult to reliably clean the images of cosmic rays (CRs). By using the \texttt{gst} catalog of resolved stars with its more stringent cuts, we avoid artifacts due to CRs. There are also problems due to charge transfer efficiency (CTE) because CTE corrections were not available for WFC3. Finally, as noted above, the images are shallow.

We have avoided some of the above problems by using the resolved stars; however, there are limitations to this choice as well. In particular, we cannot include stars below the flux limit that were not detected. Additionally, while using the \texttt{gst} catalog eliminates artifacts, it may also remove real detections. Finally, in the UVIS images, recovery of stars is biased such that artificial stars at the faint end are recovered at fainter magnitudes than those at which they are placed. We refer the reader to \citet{Williams2014a} for details on the photometric catalog. As a result, the bottom two panels in Figure \ref{fig:nuv_f275w_compare} are not complete, but they are adequate for purposes of qualitative comparison.

A detailed region-by-region comparison of the synthetic, reddened \nuv{} image and the gridded F275W image cannot be made due to differences in wavelength and the lack of completeness in the F275W stellar catalog, as discussed above. However, the two images are qualitatively very similar, and these similarities (and differences) may be important for understanding the distribution and morphology of the young, massive stars responsible for the bulk of the \nuv{} emission. Further analysis would require additional modeling, which is beyond the scope of this paper.

\section{Results}
\label{sec:results}

\begin{figure*}[]
\centering
\includegraphics[width=\textwidth]{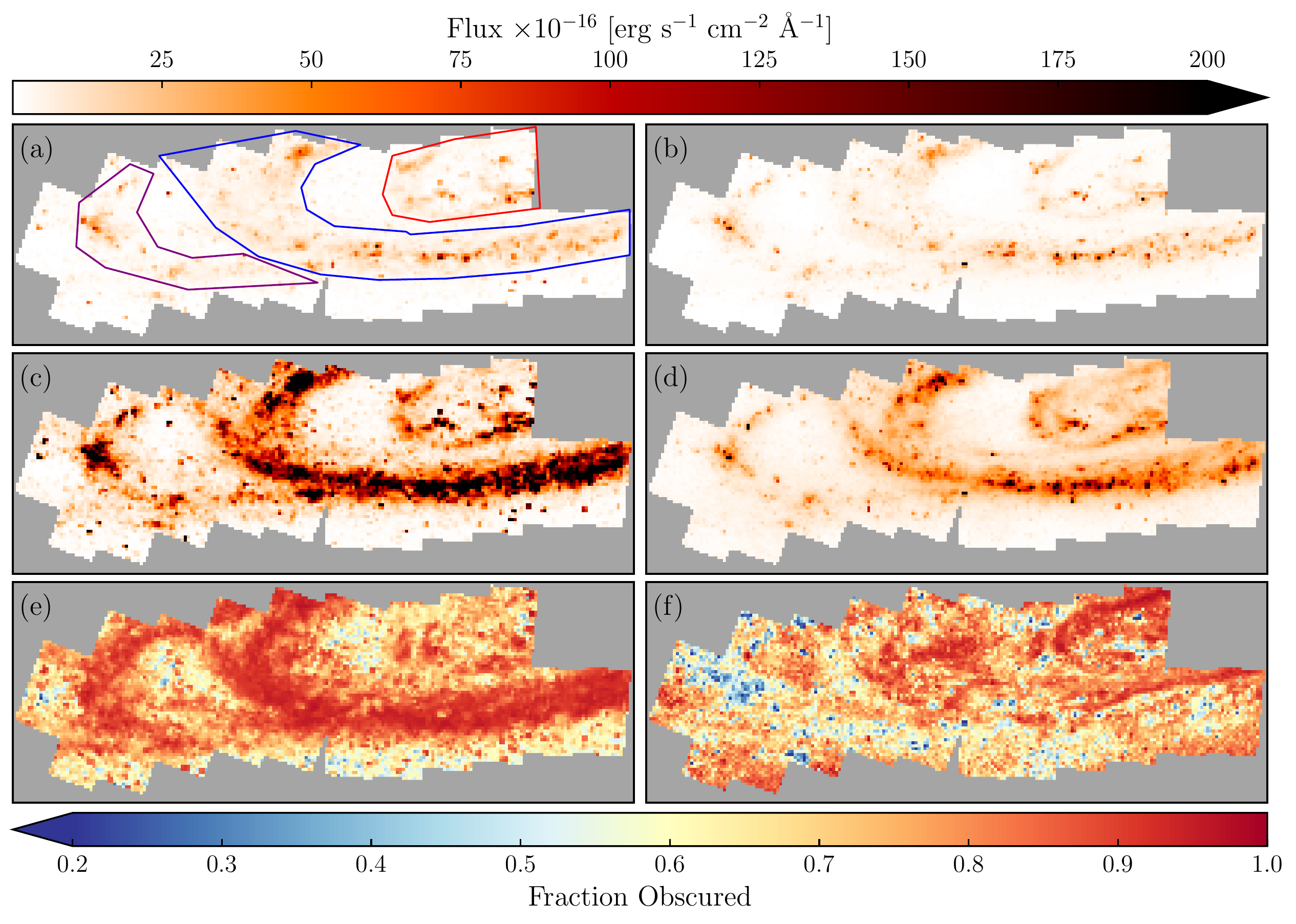}
\caption[]{Obscured Flux and Star Formation. (a) Synthetic, reddened \fuv{} flux: \ffuvsfh{}, (b) observed \galex{} \fuv{} flux: \ffuvobs{}, (c) synthetic, un-reddened \fuv{} flux: \ffuvsfhz{}, (d) observed dust-correct flux:$f_{\fuv{} + 24 \micron{}}$, (e) fraction of obscured flux from the synthetic maps: 1 - \ffuvsfh{} / \ffuvsfhz{}, (f) fraction of obscured flux from traditional observables: 1 - \ffuvobs{} / $f_{\fuv{} + 24 \micron{}}$.  Panels (a) -- (d) are on the same flux scale and panels (e) and (f) are on the same scale for ease of comparison. In panel (a), we have outlined the ring features in blue (10 kpc ring), red (inner), and purple (outer). While panels (a) and (b) show remarkable agreement, the dust-corrected flux in panels (c) and (d) looks very different. In particular, the synthetic, dust-free maps indicate much more star formation in the ring and outer arm. The difference between the dust-free maps causes the fraction of obscured flux (maps (e) and (f)) to be structurally distinct. Panel (e), which uses the synthetic data, indicates that the 10-kpc ring and the 15 kpc ring feature are heavily dust obscured, while panel (f) suggests that only parts of the 10-kpc ring are obscured but that much more of the inner ring feature is enshrouded.}
\label{fig:obscured_sf}
\end{figure*}

There are many different uses for the maps we have presented in this paper, including testing standard procedures for understanding the fraction of obscured star formation as well as the the applicability of common \sfr{} prescriptions. In this section, we compare standard corrections for dust with results from the synthetic flux to better understand \sfr{} limitations.

\subsection{Obscured Flux and Star Formation}
\label{subsec:obscured_star_formation}

In Section \ref{sec:uvmaps}, we presented synthetic \galex{} maps and verified that they were in superb agreement with observations, especially in the \fuv{} where the impact of the ``2175 \AA{} bump" is negligible. We now consider the un-reddened synthetic images (\fxsfhz) and compare them to standard estimates of the fraction of obscured star formation. We note that because \fuv{} flux maps directly to \sfr{} in a galaxy, we are using the fraction of \fuv{} flux that is obscured by dust as a proxy for the fraction of obscured star formation. We will convert to \sfr{} and discuss the effects of the obscured flux in Section \ref{subsec:sfr_measurements}.

As already discussed, the \fuv{} is an ideal tracer of star formation because it is tied directly to emission from the youngest and brightest O and B-stars. It is also highly affected by dust. Use of the \fuv{} as a monochromatic \sfr{} tracer underestimates the true \sfr{} of a galaxy. It has long been noted, though, that a correction could be made by including the flux from the longer wavelengths where dust emits. Early studies suggested using the total infrared emission

Over the past several years, it has become standard to estimate the amount of obscured star formation by ``correcting" the observed emission from young stars with the observed 24 \micron{} flux from \textit{Spitzer}. 
The idea is essentially one of energy balance. In a dust-free region, the total \sfr{} can be inferred from the observed emission from young stars. In dusty regions, some of that emission is absorbed by the dust. The total flux of light that reaches the telescope is reduced and the derived \sfr{} will be under-estimated. The dust obscures the star formation. However, the emission absorbed by the dust is not lost but rather re-emitted at different, longer wavelengths. Therefore, if we include that longer wavelength emission in our roundup of the total flux, the true \sfr{} can be accurately derived.

\citet{Calzetti2007a} first used 24 \micron{} emission as a correction to \ha{} flux. They observed a correlation between the luminosity surface densities of Pa$\alpha$ emission and 24 \micron{} emission. Pa$\alpha$ emission traces the ionizing photons from the photospheres of hot, young stars and is significantly less affected by dust than \ha{}, making it an ideal tracer of star formation in a galaxy \citep[e.g.,][]{Kennicutt1998a}. 24 \micron{} emission traces the thermal dust emission of a galaxy originating from small dust grains \citep[e.g.,][]{Draine2007a}. \citet{Calzetti2007a} combined \ha{} emission with 24 \micron{} emission and compared the result with the Pa$\alpha$ emission to derive a calibration for a \ha{} + 24 \micron{} \sfr{}. Following this result, \citet{Leroy2008a} derived a similar relation using \fuv{} emission rather than \ha{} emission. Rather than using Pa$\alpha$ emission for the total \sfr{}, they estimated the scaling factor on 24 \micron{} using various estimates of the total \sfr{}. Similarly, \citet{Hao2011a} used integrated measurements of the \citet{Moustakas2006a} sample of nearby, star-forming galaxies to compare variations of UV + monochromatic infrared luminosities with IRX (total infrared to \fuv{} luminosity ratio)-corrected \fuv{} luminosities and derive composite \sfr{} tracers, including a \fuv{} + 25 \micron{} tracer. 

We note that, in this paper, we are interested only in the 24 \micron{} correction to the \sfr{}. The idea of correcting the \fuv{} luminosity for dust using FIR emission goes back to at least the late 1990s \citep[e.g.,][]{Meurer1999a, Gordon2000a, Bell2001a}.

Dust obscuration is an extremely variable quantity. Non-active Sb-Scd galaxies typically lose about half of their bolometric luminosity to dust absorption. In early-type galaxies, this quantity drops to less than 15\% \citet{Calzetti2001b}. In M33, \citet{Boquien2015a} found 75\% of star formation in \ha{}, with only 25\% in the infrared.

In Figure \ref{fig:obscured_sf}, we combine our synthetic \fuv{} flux maps to examine the fraction of obscured star formation within the PHAT footprint and compare them with corresponding observed data from \galex{} and \textit{Spitzer}. In the left column, we show the synthetic maps derived in this paper. In the top row, we plot the synthetic, reddened flux, \ffuvsfh{}. The middle row shows the dust-corrected flux, i.e., the synthetic, dust-free flux \ffuvsfhz{}. These two maps are the same as those presented in Figure~\ref{fig:fuvmaps} but with a different scaling to emphasize the dust-free map. In the bottom row, we plot the fraction of obscured \fuv{} flux using the images in the first two rows: 1 - \ffuvsfh{} / \ffuvsfhz{}.

The increase in total flux in the ring features immediately jumps out. Most of the star formation takes place in the 10 kpc ring \citep[outlined in blue][]{Lewis2015a}, which is also the dustiest part of the galaxy. We would therefore expect to see the most change in the flux in the ring features. The width of the ring also increases, indicating that the regions on the edges of the rings are also dusty. Overall, we find that in the synthetic dust-free data there is a factor of 8 more total flux than in the synthetic reddened data. If we look at the ring features individually, the factor difference between the total flux is 7.3, 8.2, and 7.8, for the inner, 10 kpc, and outer ring features respectively. The median factor increase for all pixels in each region is 5.6, 5.1, and 4.6, respectively.  

Because M31 is a very well-studied galaxy, a lot of ancillary data exists. In addition to the \galex{} images, we also have 24 \micron{} images from \textit{Spitzer}. We can therefore compare our obscured fraction analysis with estimates from other observables. We have already compared the synthetic reddened data to the \galex{} \fuv{} image. We can also compare the synthetic dust-free data to a derived \galex{} \fuv{} + \textit{Spitzer} 24 \micron{} image. We use the prescriptions from \citet{Hao2011a} to correct the \fuv{} image for dust: 
\begin{equation}
L(\fuv{})_\textrm{corr} = L(\fuv{})_\textrm{obs} + 3.89 \times{} L(25 \micron{}).
\label{eq:24microncorrect}
\end{equation}

\noindent All luminosities have units of erg s$^{-1}$.

In the right column of Figure~\ref{fig:obscured_sf}, we present corresponding maps of traditional observables of \fuv{} + 24 \micron{} \citep[e.g.,][]{Hao2011a, Kennicutt2012a}. In the top row, we plot the observed \galex{} flux, \ffuvobs{}. The middle row shows the dust-corrected flux, i.e., the \galex{} \fuv{} flux corrected with 24 \micron{} flux, $f_{\fuv{} + 24 \micron{}}$. In the bottom row, we plot the fraction of obscured flux using the images in the first two rows. In panel (f) we use the usual observables: 1 - \ffuvobs{} / $f_{\fuv{} + 24 \micron{}}$. The 24 \micron{} data comes from \citet{Gordon2006a}.

While the synthetic reddened and the \galex{} maps look very good (total flux is conserved within 8\%), the dust-free maps are quite different. There is a factor of 2.5 more flux in the synthetic map than in the \galex{} + 24 \micron{} map. Overall, we find that 88\% of the flux is obscured in our synthetic maps, while 72\% is obscured in the observed maps. The 24 \micron{} correction therefore under-estimates the total \fuv{} flux. This has a direct effect on the quantities derived from that flux, including the \sfr{}.

We now look at the individual features of the maps. The first thing to notice is the difference in the 10 kpc ring between the two maps. If we sum up the total flux in that region, the synthetic map has a factor of 2.5 more flux than the observed map, although the median factor of the individual pixels is only 1.4. While both maps show structure within the ring, the primary difference is that the synthetic map implies that there is more flux at almost all locations. In the inner region of the galaxy, the overall structure is very similar in both maps, but the synthetic map contains a factor of 2.6 more flux, with a median of 1.9. Finally, the outer ring feature follows the same pattern. The overall structure is very similar, but the total flux is 2.4 times higher in the synthetic map with a median factor of 1.2. These differences show that in star-forming, dusty regions, the synthetic dust-free map reveals more flux than the 24 \micron-corrected \fuv{} map.

While the interarm regions lack the coherent structure of the ring features, the flux comparison in the synthetic and observed maps shows a similar trend. The total flux in the interarm regions of the synthetic map is a factor of 2.5 times higher than that in the observed map with a median value of 1.3.

The differences discussed above lead directly to the distinction in the maps of the fraction of obscured flux in the bottom row of Figure \ref{fig:obscured_sf}. As already discussed, the synthetic maps require a high degree of obscuration. Most of this dust occurs in the ring features, which are strong features in the obscured fraction map. However, all pixels have dust and are therefore obscured. The minimum and maximum obscuration in the synthetic maps is 19\% and 97\%, respectively. 

There is much less structure in the 24 \micron{}-based obscured fraction map. In this map, the most obscured regions are towards the center of the galaxy. The 10 kpc ring and outer ring feature are less apparent while the inner ring feature is stronger. Additionally, the inner edge of each ring feature is more obscured than the feature itself; i.e., it is still possible to pick out the 10 kpc ring in the 24 \micron{} obscured fraction map, though only at one edge. Finally, we note that the bright star-forming region in B21 (the bright feature at left edge of panel c) has very little obscuration in the observed data suggesting that the stars in this region are old enough to have dispersed their natal cloud but have not yet reached more evolved, dusty stages. Our results show that the 24 \micron{} correction for dust does not work well in regions of low star formation intensity which are likely dominated by older stellar populations.

\subsection{SFR Measurements}
\label{subsec:sfr_measurements}

The discrepancy between the synthetic dust-free flux map and the \fuv{} + 24 \micron{} flux map indicates that the overall \sfr{} will also be affected. The usual method of converting a \fuv{} flux into a \sfr{} is to apply an extinction correction to the observed flux, convert the flux to luminosity, and then calculate a \sfr{} according to a standard calibration. A common calibration is that from \citet{Kennicutt1998a} with updates by \citet{Hao2011a} and \citet{Murphy2011a}:
\begin{equation}
	\sfr{}= 10^{-43.35} \times L_{\fuv}.
	\label{eq:sfr}
\end{equation}

\noindent where $L_{\fuv}$ is the dust-corrected \fuv{} luminosity with units of erg s$^{-1}$ and the resulting \sfr{} has units of \sfrunit{}.

There are a variety of methods for correcting the data for dust. As mentioned in Section \ref{subsec:obscured_star_formation}, a popular method is to multiply the 24 \micron{} flux by a constant, $w$, especially in relatively nearby galaxies where high quality \textit{Spitzer} 24 \micron{} data exists. Another method is to use UV color ($m_{\fuv{}} - m_{\nuv{}}$) to correct the \fuv{} magnitude which is then converted to flux for further conversion to \sfr{}. This method is often used at high redshift.

We use the calibrations from \citet{Hao2011a}. The correction using 24 \micron{} data is shown in Equation \ref{eq:24microncorrect}. The UV color correction is:
\begin{equation}
\begin{split}
	A_{\textrm{FUV}} = \\
	(3.83 \pm & 0.48) [(\fuv{} - \nuv)_{\textrm{obs}} - (0.022 \pm 0.024)].
\end{split}
\end{equation} 

\begin{figure*}[]
\centering
\includegraphics[width=0.95\textwidth]{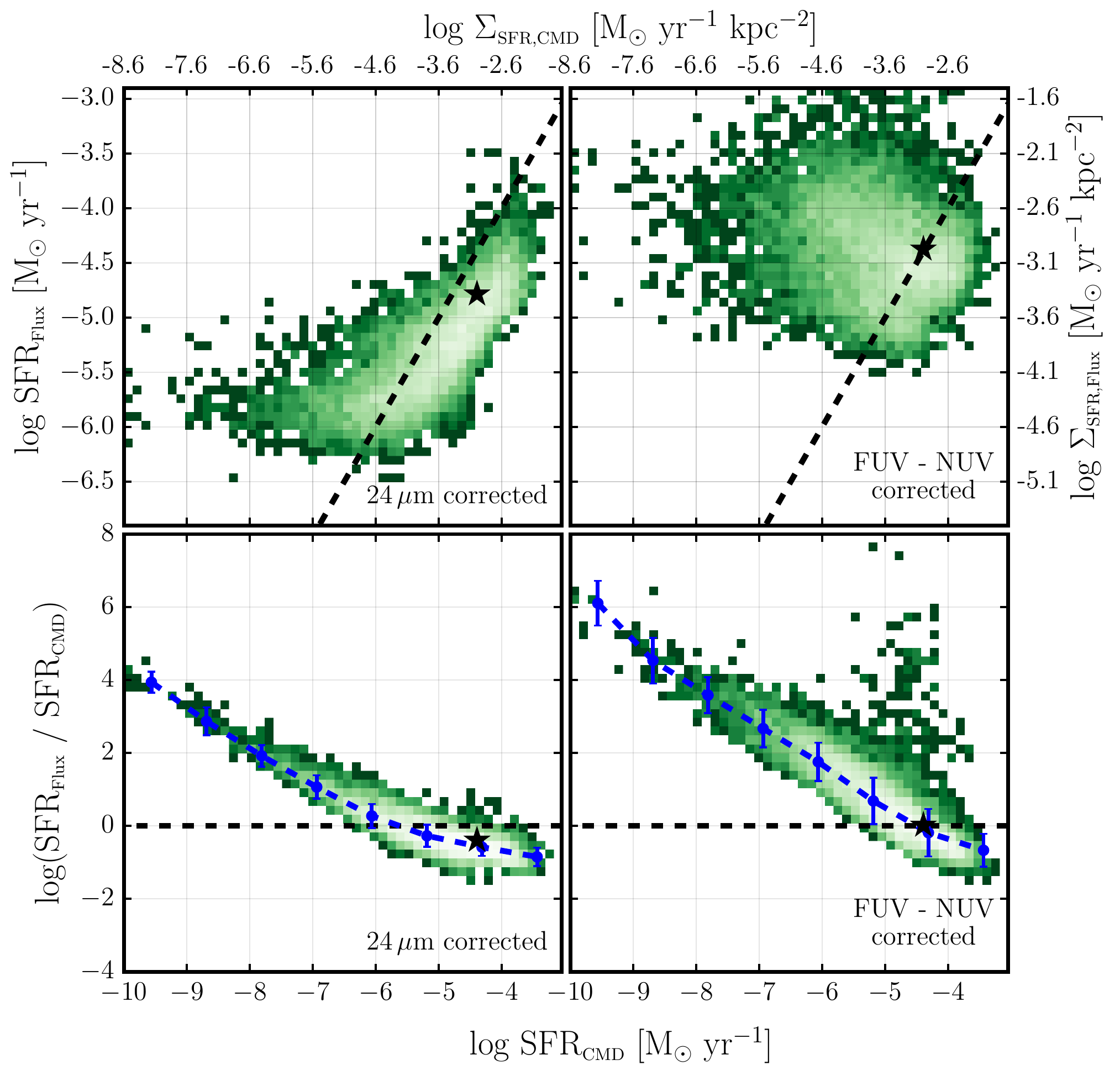}
\caption[]{Comparison between CMD-based and flux-based \sfr{s}. In all panels, the x-axis is the \sfr{} from the CMD-derived \sfh{s}, averaged over the last 100 Myr (\sfroneh{}). In the top left panel, the y-axis shows the flux-based \sfr{} derived from the \fuv{} + 24 \micron{} combination. The top right panel is corrected for dust using \fuv{}--\nuv{} color. The bottom panels show the ratio of the flux-based \sfr{} to the CMD-based \sfr{}. In each panel, the black star marks the flux-weighted mean along each axis, and the black dashed line denotes one-to-one agreement.}
\label{fig:sfr_compare_dust_correct}
\end{figure*}

In Figure~\ref{fig:sfr_compare_dust_correct}, we examine the relationship between the \sfr{} averaged over the last 100 Myr as calculated from the CMD-derived \sfh{} \citep{Lewis2015a} and that derived from \fuv{} flux corrected for extinction with the 24 \micron{} or UV color calibrations described above. In each panel, we plot the CMD-derived \sfr{} ($\sfr{}_\textrm{CMD}$) on the x-axis. The top panels show the flux-to-flux relationship. In the bottom panels, we plot the ratio of the flux-based \sfr{} to the cmd-based \sfr{} ($\sfr{}_\textrm{flux} / \sfr{}_\textrm{CMD}$) on the y-axis. In each panel, the black dashed line represents one-to-one agreement. In the top panels, the black star marks the flux-weighted mean along each axis. In the bottom panels, we have plotted the running median and standard deviation in blue.

In the left panels, we show the flux-based \sfr{} ($\sfr{}_\textrm{flux}$) derived after using the 24 \micron{} prescription to correct for extinction. In the right panels, we show the flux-based \sfr{} calculated from \fuv{} flux corrected for dust with the UV color calibration. 

We first look at the left panels. On the top we plot the \sfr{} derived from the 24 \micron{}-corrected \fuv{} flux against the CMD-derived \sfr{} averaged over the most recent 100 Myr. The overall morphology of the data is ok using the 24 \micron{} correction. There is a clear trend in the data that $\sfr{}_\textrm{flux,24\micron}$ increases as $\sfr{}_\textrm{CMD}$ increases, as is expected. The flattening in $\sfr{}_\textrm{flux,24\micron}$ occurs at $\sfr{}_\textrm{CMD} < \about10^{-6}$\, which corresponds approximately to the limit at which the flux calibrations are no longer reliable, especially on these spatial scales \citep[e.g.,][]{Murphy2011a, Leroy2012a, Kennicutt2012a}. This results in an over-estimate of the \sfr{} at low  $\sfr{}_\textrm{CMD}$ and an under-estimate at high $\sfr{}_\textrm{CMD}$. This trend is seen more clearly in the bottom left panel, where we plot the $\log{}$ of the ratio between  $\sfr{}_\textrm{flux,24\micron}$ and  $\sfr{}_\textrm{CMD}$. The over-estimate of  $\sfr{}_\textrm{flux,24\micron}$ increases as  $\sfr{}_\textrm{CMD}$ decreases. At higher  $\sfr{}_\textrm{CMD}$, the ratio starts to flatten though offset from the one-to-one line.

While the overall morphology of the relationship between $\sfr{}_\textrm{flux}$  and $\sfr{}_\textrm{CMD}$ is good, the flux-weighted mean, marked by the black star in the top left panel, is offset. We find that the mean flux-based \sfr{} is 0.39 dex lower than the mean CMD-based \sfr{}. The 24 \micron{}-corrected \fuv{} flux will therefore under-estimate the \sfr{} by a factor of \about2.5. 

The right panels in Figure \ref{fig:sfr_compare_dust_correct} show the comparison between $\sfr{}_\textrm{CMD}$ and $\sfr{}_\textrm{flux,\fuv-\nuv}$. The results are very different from the 24 \micron{} correction. When using UV color, the morphology of the relationship is completely wrong. On a region-to-region basis, there is zero correlation between $\sfr{}_\textrm{CMD}$ and $\sfr{}_\textrm{flux,\fuv-\nuv}$. Any $\sfr{}_\textrm{CMD}$ can map to a two order of magnitude range in $\sfr{}_\textrm{flux,\fuv-\nuv}$. The only trend that exists is that the UV color correction results in an over-estimate of $\sfr{}_\textrm{flux,\fuv-\nuv}$ at low $\sfr{}_\textrm{CMD}$ and an under-estimate at high $\sfr{}_\textrm{CMD}$. Despite the absence of region-to-region correlation, the flux-weighted mean \sfr{} falls directly on the one-to-one line. 

We note that we could have used the synthetic, dust-free \fuv{} flux-derived \sfr{} on the x-axis instead of the 100 Myr averaged \sfr{} from the CMD-derived \sfh{s}. The results would be the same with an offset of the mean flux-weighted \sfr{s} from the one-to-one line when using the 24 \micron{} correction and a lack of correlation when using the UV color correction.

\section{Discussion}
\label{sec:discussion}

We have shown that the conversion of \fuv{} flux into a \sfr{} after correcting for dust with two different prescriptions results in vastly different relations when compared with the \sfr{} determined from the CMD-derived \sfh{} (or alternatively from using Equation \ref{eq:sfr} on the dust-free synthetic \fuv{} flux, \ffuvsfhz). These differences are perhaps not surprising, but they do beg the question of what drives the discrepancies. At the most basic level, the discrepancy arises either from the modeled data (synthetic flux or CMD-based \sfr) or from the \sfr{} calibrations in the literature.
 
As discussed in \citet{Lewis2015a} and Section \ref{subsec:issues}, there are a variety of factors that affect the CMD-derived \sfh{s} and the synthetic flux. The primary uncertainties come from the chosen dust parameters and uncertainties in the \sfh{s}, which were only of order 10\% over a 100 Myr timescale. We also showed in Section \ref{sec:uvmaps} and Figure \ref{fig:fluxratiocompare} that the synthetic, reddened flux is in very good agreement with the observed \galex{} \fuv{} flux. This suggests that the dust parameters derived from the \sfh{s} are reasonable and therefore the \sfr{} derived from the CMD-derived \sfh{s} is not driving the discrepancy.

The inconsistencies must therefore lie with the flux calibrations. This should not be a surprising conclusion. The vast majority of \sfr{} prescriptions are designed for regions containing at least tens of thousands of stars, such as entire galaxies or bright, star-forming regions. They make assumptions about the shape of the \sfh{}, usually assuming a constant \sfr{} over some timescale. They assume that the IMF in the given region is completely sampled. They assume that the region is homogenous, i.e., stellar populations are completely mixed and they are insensitive to structure in the dust, gas, and stars. Additionally, they often use only a single metallicity (usually) solar, and they are insensitive to metallicity evolution. 

Most objects in the universe fall into the unresolved regime where the above assumptions are generally safe. This is especially true in the high redshift universe. In this regime, the UV color correction is commonly used to correct the \sfr{} for dust. While Figure \ref{fig:sfr_compare_dust_correct} suggests that this correction is terrible in the spatially-resolved regime, the fact that the flux-averaged mean \sfr{s} are in agreement suggests that it may be a reasonable choice in the unresolved regime, especially when the availability of data at IR wavelengths is limited. 

In the local Universe,the existing calibrations seem to work on large ($>$1 kpc) scales, which is the scale at which they were calibrated. However, we showed in this paper that even the flux-weighted means over the entire galaxy are off when using the 24 \micron{} correction. Additionally, even on these scales, many of the above assumptions are not applicable. Nearby galaxies show highly-defined structure in spiral arms and rings, in dust lanes, and in populations of molecular clouds and other gaseous objects. This structure varies in population; spiral arms are populated by young and blue, main sequence stars. The centers of galaxies tend to contain older, more redder populations. Additionally, as structure within a galaxy is resolved, variations in stellar environment become apparent. Regions of high and low star formation, of varying metallicity, and of changing stellar density are revealed. Finally, the shape of the \sfh{} is important, especially when correcting for dust \citep[e.g.,][]{Boquien2016a}. The assumptions made when averaging over entire galaxies can no longer be made on smaller, variable scales. In fact, making such assumptions will bias the results. Nonetheless, we still want to determine \sfr{s} for these galaxies and on these scales to take advantage of the increased spatial resolution. 

In this paper, we have chosen to use a single prescription for each method of correcting \fuv{} flux for dust. However, many studies have examined hybrid \sfr{} tracers in an effort to determine the best combination of the data. Specifically, looking at the \fuv{} + 24 \micron{} combination, the uncertainty primarily lies on the weight applied to the 24 \micron{} data, $w_{24}$. In this paper, we use the \citet{Hao2011a} factor of 3.89, which was derived for a sample of 133 nearby, star-forming galaxies. \citet{Zhu2008a} also derived a calibration between \fuv{} and 24 \micron{} and found $w_{24}=6.31$. Both of these measurements were derived on global scales with a single data point for each galaxy.

Some work has been done on sub-galactic scales. \citet{Leroy2008a} presented a calibration for the  \fuv{} + 24 \micron{}-based \sfr{} based on a sample of 23 nearby galaxies at 750 pc scales. Modulo a typo in their appendix D, their calibration requires that $w_{24}=6.0$ \citep{Liu2011a}. The \citet{Leroy2008a} value of $w_{24}$ is similar to that of \citet{Zhu2008a}, with the notable difference in the physical scales on which it was derived (i.e., 750 pc vs. entire galaxies). As discussed above, using relations on scales other than which they were derived can be precarious. In light of this, it would seem to be safer to use the \citet{Leroy2008a} calibration instead of the \citet{Hao2011a} calibration.

However, using $w_{24} = 6.0$ does not significantly change the results presented in this study. The primary reason for this lack of change is that the calibration derived by \citet{Leroy2008a} and stated in \citet{Liu2011a} differs by more than just the value of $w_{24}$. The \citet{Hao2011a} prescription for \sfr{} derived from \fuv{} + 24 \micron{} observations is:
\begin{equation}
	\sfr{}= 4.47 \times 10^{-44} \; (L_{\fuv} + 3.89 \; L_{24\small\micron{}}),
\end{equation}
which is simply a combination of Equations \ref{eq:24microncorrect} and \ref{eq:sfr}. The \citet{Leroy2008a} prescription is:
\begin{equation}
	\sfr{}= 3.40 \times 10^{-44} \; (L_{\fuv} + 6.0 \; L_{24\small\micron{}}).
\end{equation}
The difference in the conversion factor from luminosity to \sfr{} effectively offsets the differences that result from the change in the value of $w_{24}$. Therefore, we have chosen not to present new plots created by using this slightly different, spatially-resolved calibration factor, but instead describe the changes that result.

A change in the \sfr{} calibration will have an effect on the fraction of obscured flux (Figure \ref{fig:obscured_sf} and on the relationship between the flux-based \sfr{} and the CMD-based \sfr{} (Figure \ref{fig:sfr_compare_dust_correct}). In Figure \ref{fig:obscured_sf}, the primary difference is that the regions of highest flux shown in panel d have slightly more flux when using the \citet{Leroy2008a} calibration, and the fraction of obscured star formation (panel e) is therefore higher across the galaxy. The total obscured flux increases from 72\% to 80\%, the overall morphology of the maps does not change, and the differences discussed in Section \ref{subsec:obscured_star_formation} remain.

Increasing $w_{24}$ improves the relationship between the CMD-based \sfr{} and the flux-based \sfr{} improves; however, as already noted, that improvement is offset by the change in the multiplicative factor that converts a dust-corrected luminosity to a \sfr{}. The \sfr{} derived with the \citet{Leroy2008a} prescription is offset from the CMD-based \sfr{} by a factor of \about2.3 compared to \about2.5 for the \citet{Hao2011a} calibration. Consequently, Figure \ref{fig:sfr_compare_dust_correct} is largely unchanged. Therefore, while it is generally acknowledged that the global calibration of $w_{24}$ will break down on small scales, the local calibration -- derived on scales that are 2--8 times larger than those presented in this study --  does not perform much better.

More recently, \citet{Boquien2016a} calculated $w_{24}$ (they call it $k_i$ and examine other infrared tracers longward of 24 \micron{} as well) in 8 different galaxies and found a wide distribution ranging from 1.55 to 13.45 with a mean of 8.11. They stress that this factor has important galaxy-to-galaxy variations, as well as local vs. global variations that should not be overlooked. Uncertainty in this factor is is clearly an issue in this study as well. Had we used the \citet{Boquien2016a} value of 8.11 in this paper, we would have found much better agreement between the flux-based \sfr{} and the CMD-based \sfr{}.

\citet{Boquien2016a} actually present two different parameterizations of their factor $k_i$ based on \fuv-NIR color and on NIR luminosity density. These parameterizations allow $k_i$ to vary from galaxy-to-galaxy. However, they emphasize that these parameterizations should not be used on scales below 500 pc, and therefore should not be used at the spatial scales presented here. Further testing and analysis of these parameterizations is beyond the scope of this paper. It is clear, though, that as higher spatial resolution becomes available beyond the most nearby galaxies, such methods for reliably determining the \sfr{} while accounting for galaxy-to-galaxy variation are needed.

\section{Conclusions}
\label{sec:conclusions}

We have used spatially-resolved \sfh{s} derived from optical resolved star data to model the SEDs of over 9000 sub-kpc regions in M31 and produce detailed maps of synthetic reddened and dust-free UV flux across the entire area covered by the PHAT survey. The \sfh{s} were derived by \citet{Lewis2015a}\ using \acsb{} and \acsi{} photometry from the PHAT survey. Both intrinsic and attenuated SEDs were derived from the \sfh{s} using FSPS. These SEDs were convolved with the \galex{} \fuv{} and \nuv{} response curves to generate synthetic fluxes, \fxsfh{}. All of the flux values were then assembled into an overall map using Montage. The pixels correspond to physical areas of $4.4\times 10^4\pc^2$. 

We also used Montage to construct maps of the observed UV flux, \fxobs{}, using \galex{} DIS images and 24 \micron{} \citep{Gordon2006a}, as well as \sfroneh{} from the spatially-resolved \sfh{s}, all on the same spatial scale. 

The agreement between the observed flux maps and the reddened modeled flux maps is encouraging, especially given that they were derived from only photometry in two optical bands. They indicate that models of UV emission and dust extinction are fairly accurate. The median log ratio of the synthetic reddened flux to the observed flux is 0.03 and -0.03 in the FUV and NUV, respectively, with standard deviations of 0.24 in the \fuv{} and 0.16 in the \nuv{}. This agreement confirms the robustness of our modeling procedure and justifies the assumptions made in the modeling routine.

The scatter in the relation between observed \galex{} flux and synthetic, reddened flux is primarily due to uncertainties on the \sfh{s} used in the modeling process. These uncertainties include a factor due to incomplete sampling of the IMF. 

We compared our synthetic reddened and dust-free \fuv{} flux maps with corresponding maps from observations. The synthetic reddened map was compared to the observed \galex{} \fuv{} map. The dust-free map was compared to map derived from a combination of \fuv{} + 24 \micron{} data. In the synthetic \fuv{} flux maps, 88\% of the flux is obscured by dust. In the observed maps, 71\% is obscured. This suggests that the 24 \micron{} correction results in an under-estimate of the total \fuv{} flux.

We converted the observed flux map into a map of \sfr{} using two common prescriptions, one using 24 \micron{} to correct for dust and the other using UV color. We compared the resulting \sfr{s} with those determined from the \citet{Lewis2015a} \sfh{s} averaged over the past 100 Myr. While the morphology of the relation between the CMD-derived \sfr{} and that calculated using the 24 \micron{} correction was good, the \galex{} + 24 \micron{}-derived \sfr{} was under-estimated by a factor of 2.5. Conversely, the UV-color correction resulted in a flux-weighted mean \sfr{} in good agreement with that derived from the CMDs, but the relation shows no correlation between the two \sfr{s} as would be expected. 

The results we present in this paper provide an end-to-end verification of the \sfh{} results presented in \citet{Lewis2015a} and show that we have the ability to model flux in a variety of bandpasses given optical data and appropriate assumptions for the stellar IMF, a set of models describing stellar spectra and evolution, and an extinction model. While modeling flux in other bands provides different challenges, the technique is extremely promising.

\acknowledgements
The authors would like to thank the anonymous referee for thoughtful comments that improved the clarity of this paper. This research has made use of NASA's Astrophysics Data System Bibliographic Services and the NASA/IPAC Extragalactic Database (NED), which is operated by the Jet Propulsion Laboratory, California Institute of Technology, under contract with the National Aeronautics and Space Administration. This work was supported by the Space Telescope Science Institute through GO-12055. This research made use of Astropy, a community-developed core Python package for Astronomy \citep{Astropy2013a}, as well as NumPy and SciPy \citep{Oliphant2007a}, IPython \citep{Perez2007a}, and Matplotlib \citep{Hunter2007a}. This research made use of Montage. It is funded by the National Science Foundation under Grant Number ACI-1440620, and was previously funded by the National Aeronautics and Space Administration's Earth Science Technology Office, Computation Technologies Project, under Cooperative Agreement Number NCC5-626 between NASA and the California Institute of Technology.

\appendix

\section{The Effects of Old Stellar Populations on Synthetic and Observed Flux}
\label{app:oldpops}

Throughout the main body of this paper, we analyzed synthetic and observed \fuv{}, \nuv{}, and 24 \micron{} flux without regard to the stellar populations emitting at each wavelength. We used the entire \sfh{} -- from the present day to 14 Gyr ago -- in the modeling routine (Section \ref{subsec:flux_modeling}, despite the warning that the \citet{Lewis2015a} \sfh{s} are robust only to 500 Myr ago). We did this in order to accurately compared the modeled UV flux with the total observed \galex{} flux. While most of the UV flux comes from stars that are 100 to 300 Myr, approximately 20-30\% of the emission at \fuv{} and \nuv{} wavelengths comes from stars that are older than this (see discussion in Section \ref{subsec:emissiontimescales}).

We must include these older populations to accurately model the UV flux. We could have made an assumption about the form of the \sfh{} beyond 500 Myr (constant, declining tau, etc.); however, we decided to use the full \sfh{s} derived from the CMD analysis. We note that while the \sfh{} solution was optimized for the most recent 500 Myr, it still has information at older ages. Using this total \sfh{} will be more robust than using an assumed form.

\subsection{Modeled-to-Observed Flux Comparison}

\begin{figure}[]
\centering
\includegraphics[width=\textwidth]{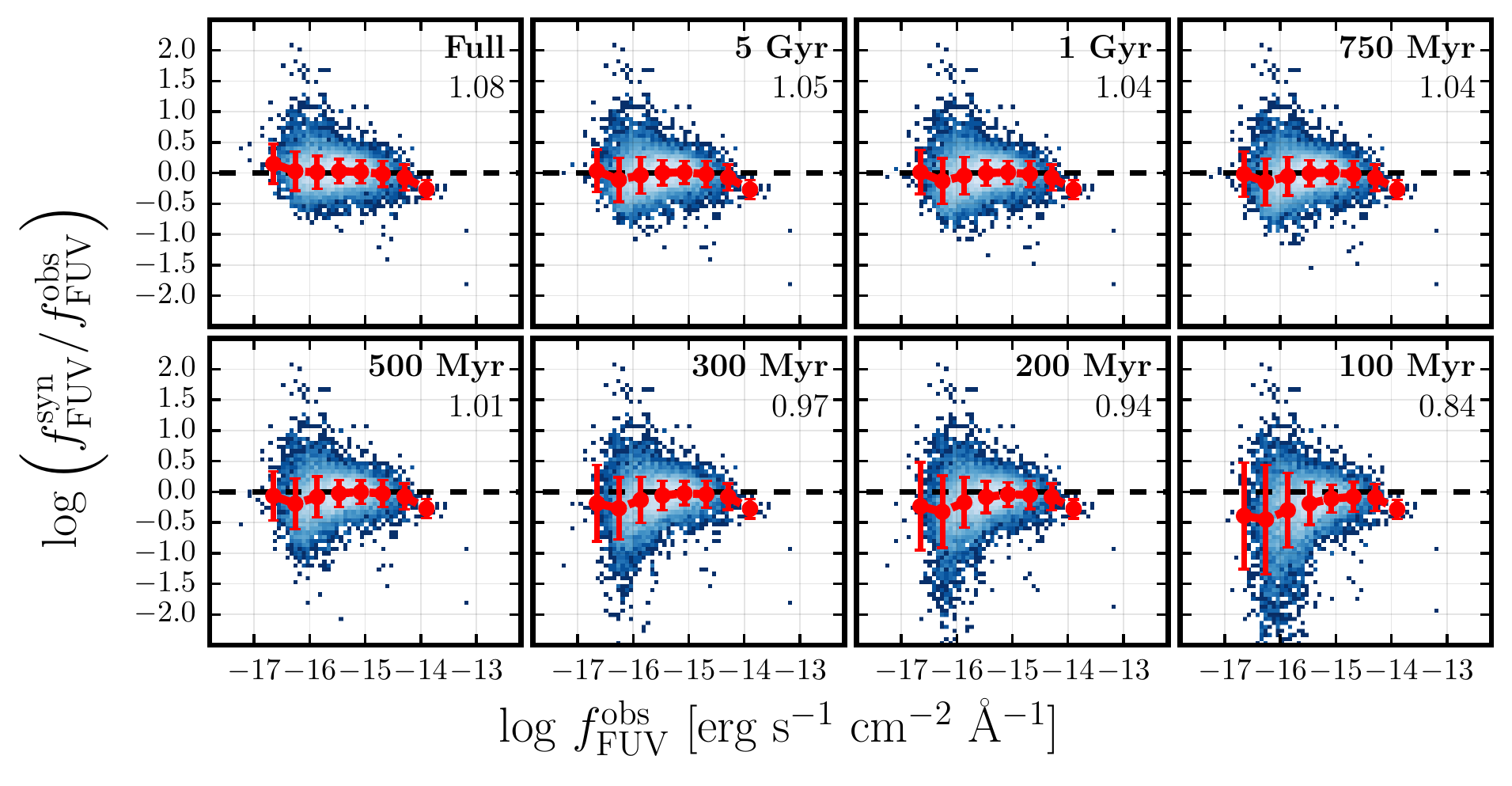}
\caption[]{The log ratio of the synthetic, reddened \fuv{} flux to the observed \galex{} \fuv{} flux as a function of observed flux with a variety of age cuts on the \sfh{}. The red circles in each panel show the running median with the standard deviation given by the error bars on each point. The age cut used is listed in the top right corner of each panel. The fraction of total synthetic flux to total observed flux is listed just below the age cut.}
\label{fig:flux_compare_fuv_no_old_correct}
\end{figure}

We account for these older populations in both the synthetic and the observed data. In Figure \ref{fig:flux_compare_fuv_no_old_correct}, we examine the effect of older stellar populations on the synthetic flux. We plot the ratio of the synthetic flux to the observed flux as a function of the observed flux (as in Figure \ref{fig:fluxratiocompare}). In each panel, we have modeled the \fuv{} flux with a different age limit applied to the \sfh{}. The top left panel uses the full \sfh{} and is identical to the top panel in Figure \ref{fig:fluxratiocompare}. In the next panel we model the flux using the \sfh{} back to 5 Gyr ago. The bottom left panel applies a cut to the \sfh{} at 500 Myr ago, which is the limit specified in \citet{Lewis2015a}. The bottom right panel uses \sfr{} data from just the most recent 100 Myr. In each panel, we have indicated the age limit in the top right corner. Below that is the ratio of the sum of the synthetic flux to the sum of the observed flux. The red circles and error bars show the running median and standard deviation. 

The effect of reducing the age information in the flux modeling routine is clear. As the information from older populations is removed, the total synthetic flux drops, as expected. This decrease is generally seen in the low flux regions. The total synthetic flux decreases by \about20\% as the age information is cut from 14 Gyr to 100 yr. This is in line with literature values of the amount of UV flux expected from older populations \citep[e.g.,][]{Johnson2013a}. We note, however, that in all panels and all flux bins, the ratios are consistent with zero to within $1\sigma$. 

\begin{figure}[]
\centering
\includegraphics[width=\textwidth]{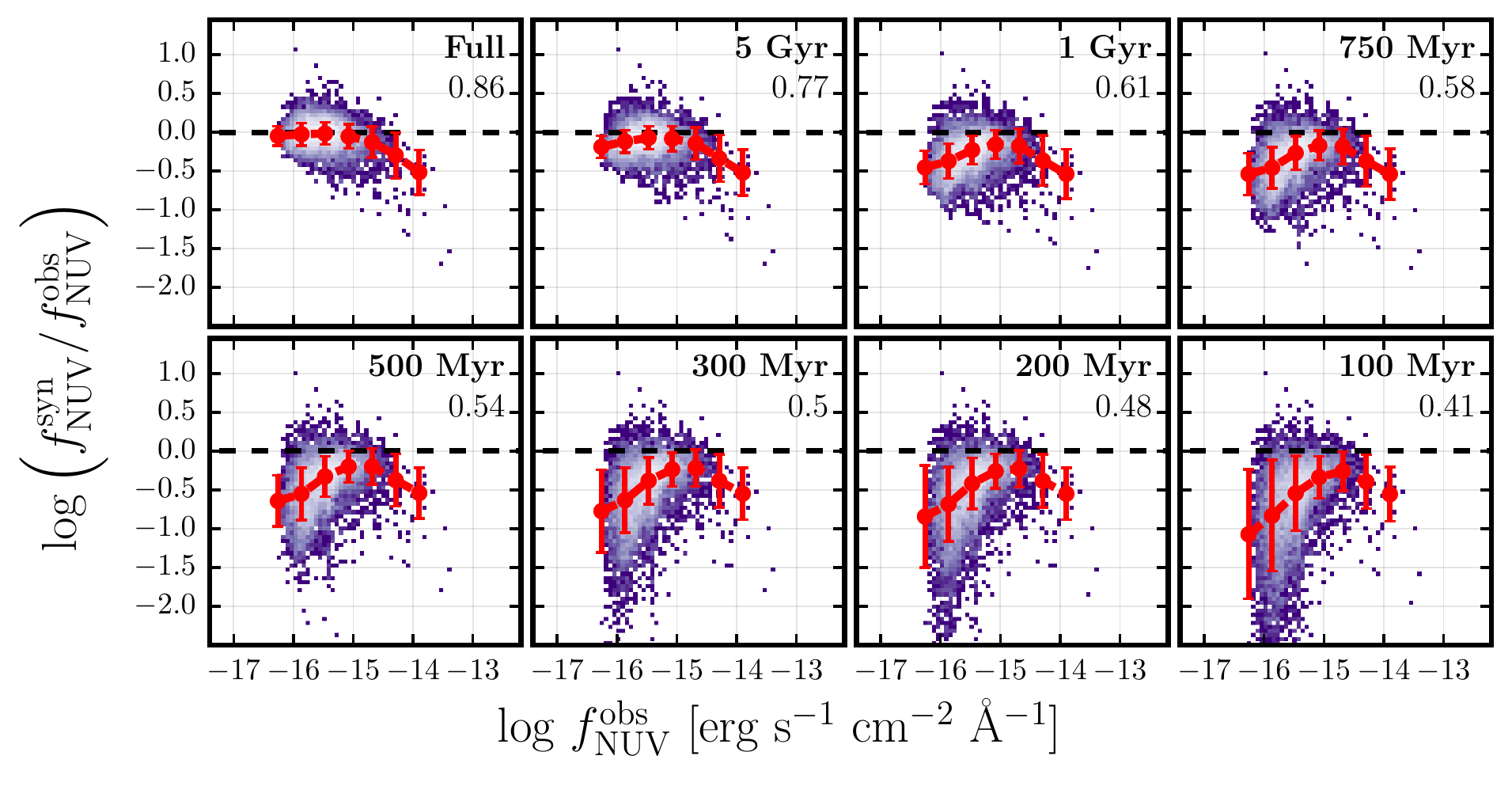}
\caption[]{The same as Figure \ref{fig:flux_compare_fuv_no_old_correct} except for the \nuv{}.}
\label{fig:flux_compare_nuv_no_old_correct}
\end{figure}

We perform the same exercise in the \nuv{} and plot the results in Figure \ref{fig:flux_compare_nuv_no_old_correct}. The general results are similar to those in the \fuv{} except the decrease in flux is much more substantial in the \nuv{}, especially at the low-flux end. As mentioned in our discussion of Figure \ref{fig:fluxratiocompare}, the \nuv{} is much more sensitive to the shape of the extinction curve, primarily at the high-flux end. This sensitivity likely contributes to the under-production of synthetic flux seen in the top left panel when the full \sfh{} is used and propagated through as the synthetic flux relies more strongly on younger populations that may be more affected by the 2175\AA{} bump.

The observed data is also affected by emission from older stellar populations \citep[e.g.,][]{Kennicutt2009a}. A few corrections exist in the literature. One such method is to use 3.6 \micron{} flux to correct for older populations. Some fraction of the 3.6 micron data, $\alpha_{3.6, \fuv}$, is subtracted from the \fuv{} data:
\begin{equation}
	I_{\fuv,{\rm young}} = I_{\fuv,{\rm all}} - \alpha_{3.6,\fuv} \times I_{3.6\micron}
\end{equation}
\noindent leaving the \fuv{} intensity from young stars. The general idea is that older stars tend to be faint at bluer wavelengths but will emit more strongly in the infrared. \citet{Leroy2008a} looked at the ratio of \fuv{} to 3.6 \micron{} \sfr{} intensities in their sample of nearby galaxies and found $\alpha_{3.6,\fuv} \sim 2-4\times10^{-3}$. However, \citet{Ford2013a} performed the same exercise in M31 and found $\alpha_{3.6,\fuv} \sim 8\times10^{-4}$. Here we use the \citet{Ford2013a} value derived for M31.

\begin{figure*}[]
\centering
\includegraphics[width=\textwidth]{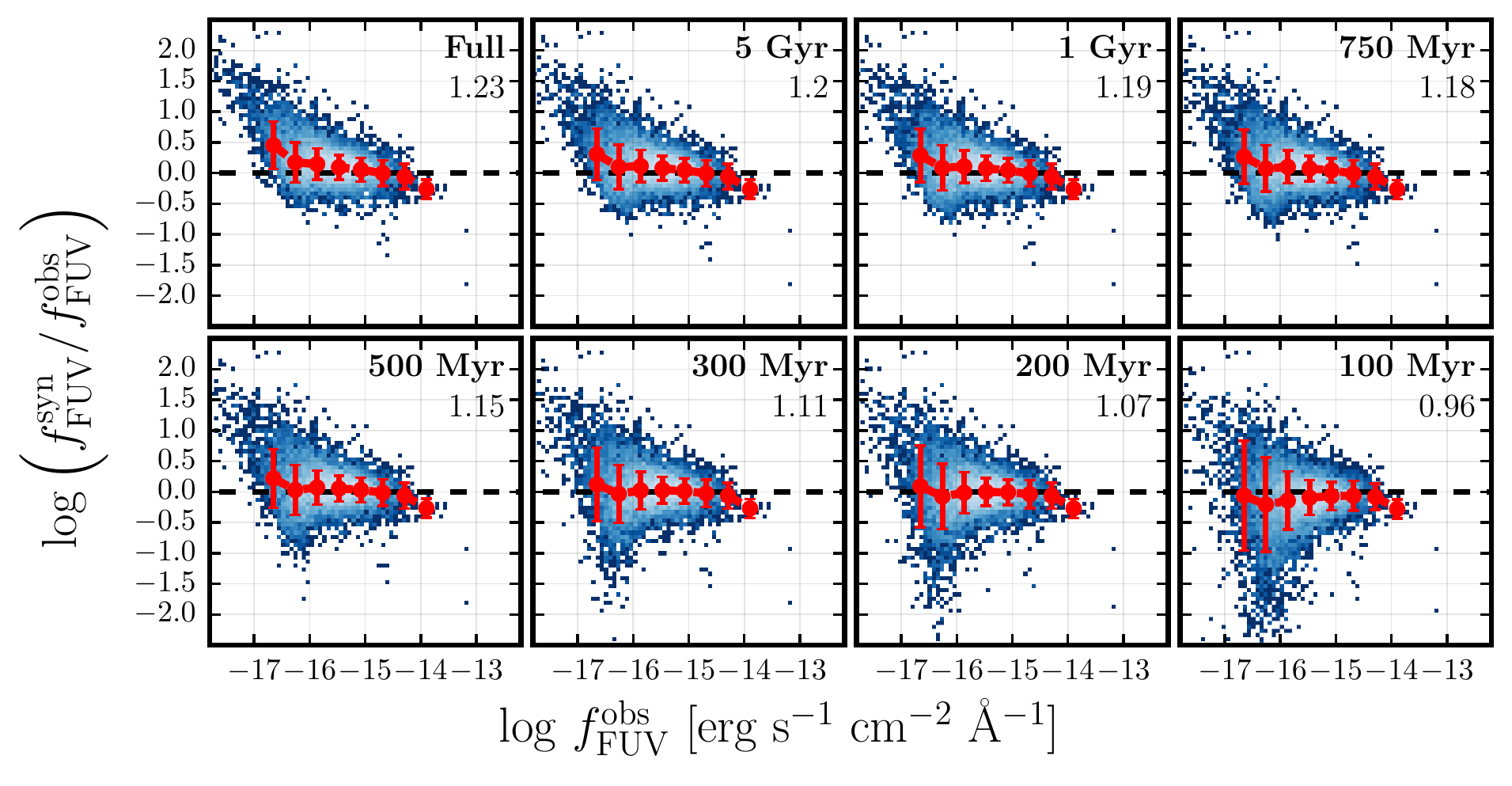}
\caption[]{The same as Figure \ref{fig:flux_compare_fuv_no_old_correct} except that we also include a correction for older populations in the observed flux.}
\label{fig:flux_compare_fuv_old_correct}
\end{figure*}

We apply this correction to the observed \fuv{} flux and we plot the resulting flux ratios in Figure \ref{fig:flux_compare_fuv_old_correct}. We have also included the same age limit cuts to the synthetic flux as we showed in Figure \ref{fig:flux_compare_fuv_no_old_correct}. The correction to the observed flux increases the flux ratios, especially at the low-flux end, as would be expected. As the age limit on the synthetic flux decreases, the correction to the observed flux has less of an impact. In all panels and all flux bins (except for the brightest bin), the synthetic and observed fluxes are in agreement within $1\sigma$. There is no corresponding correction in the literature for the contribution of older populations to the \nuv{} flux.

\subsection{SFR Comparison}

In this section, we examine how the effects of older stellar populations on the observed flux translate to changes in the derived \sfr{}. In the previous section, we corrected the observed \fuv{} flux for older stellar populations by using data at 3.6 \micron{}. We can do the same thing with our 24 \micron{} data:

\begin{equation}
	I_{24,{\rm young}} = I_{24,{\rm all}} - \alpha_{3.6,24} \times I_{3.6\micron}
\end{equation}

\citet{Leroy2008a} found $\alpha_{3.6, 24} \sim 0.1$ in their sample of galaxies. In their study of M31, \citet{Ford2013a} also found $\alpha_{3.6, 24} = 0.1$, so we use that value here.

\begin{figure*}[h]
\centering
\includegraphics[scale=0.75]{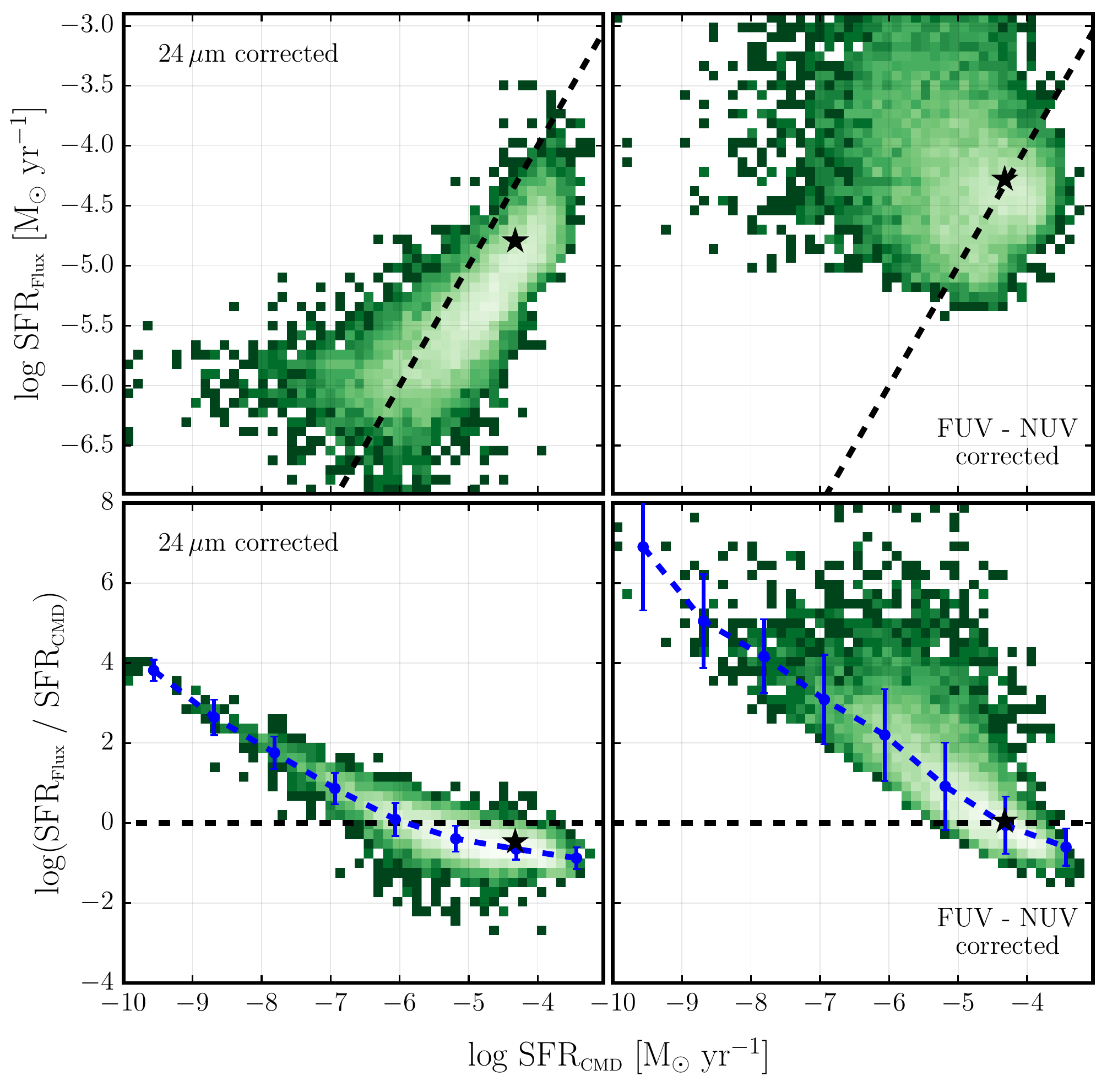}
\caption[]{Comparison between CMD-based and flux-based \sfr{s}. This is the same as Figure \ref{fig:sfr_compare_dust_correct} except that we have corrected the observed \fuv{} and 24 \micron{} fluxes for the contribution from older stellar populations. In all panels, the x-axis is the \sfr{} from the CMD-derived \sfh{s}, averaged over the last 100 Myr (\sfroneh{}). In the top left panel, the y-axis shows the flux-based \sfr{} derived from the \fuv{} + 24 \micron{} combination. The top right panel is corrected for dust using \fuv{}--\nuv{} color. The bottom panels show the ratio of the flux-based \sfr{} to the CMD-based \sfr{}. In each panel, the black star marks the flux-weighted mean along each axis, and the black dashed line denotes one-to-one agreement.}
\label{fig:sfr_compare_fuv_old_correct}
\end{figure*}

We corrected both the \galex{} \fuv{} and the \textit{Spitzer} 24 \micron{} data for contributions from older stellar populations and recalcualted the \sfr{s}. We plot the resulting \sfr{s} compared with those derived from the CMD-derived \sfh{} in Figure \ref{fig:sfr_compare_fuv_old_correct}. This figure can be directly compared to Figure \ref{fig:sfr_compare_dust_correct} which does not include the old star correction.

Including the correction for older stars worsens the agreement between the observed and CMD-based \sfr{s}. 
In Figure \ref{fig:sfr_compare_fuv_old_correct}, the x-axis remains unchanged from that in Figure \ref{fig:sfr_compare_dust_correct} because it is taken directly from the most recent 100 Myr of the CMD-derived \sfh{s}. No correction for old stars is necessary. In the left panels, the values on the y-axis decrease because there is generally less \fuv{} flux. The overall morphology remains the same between Figures \ref{fig:sfr_compare_fuv_old_correct} and \ref{fig:sfr_compare_dust_correct}; however, the mean values of the flux and CMD-based \sfr{s} now differ by 0.46 dex (a factor of 2.9), up from 0.39 dex (a factor of 2.4).

In the right panels of Figure \ref{fig:sfr_compare_fuv_old_correct}, we show the UV color-derived \sfr{}. While the flux-weighted means along the two axes are still in agreement, there is no overall correlation between the two \sfr{s} on small spatial scales, as seen in in Figure \ref{fig:sfr_compare_dust_correct}. We cannot, however, draw any conclusions from the right panels because the \nuv{} data have not been corrected for the contribution from older stellar populations.

\mbox{~}
\clearpage

\bibliographystyle{apj}
%\bibliography{/Users/alexialewis/research/sfh}

\end{document}